\documentclass[12pt]{article}
\usepackage{amsmath,amsfonts,amssymb,graphicx,epsfig}
\usepackage[usenames]{color}
\usepackage{pst-all}
\numberwithin{equation}{section}

\newcommand{\nc}{\newcommand}
\nc{\fh}{\hat{f}}
\nc{\muh}{\hat{\mu}}
\nc{\nuh}{\hat{\nu}}
\nc{\bib}{\bibitem}
\nc{\al}{\alpha}
\nc{\g}{\gamma}
\nc{\G}{\Gamma}
\nc{\D}{\Delta}
\nc{\eps}{\epsilon}
\nc{\la}{\lambda}
\nc{\La}{\Lambda}
\nc{\var}{\varphi}
\nc{\pa}{\partial}
\nc{\nn}{\nonumber \\ }
\nc{\hf}{\frac{1}{2}}
\nc{\dz}{\frac{dz}{2\pi i}}
\nc{\bin}[2]{\left(\!\!\!\begin{array}{c} {#1}\\ {#2} \end{array}\!\!\!\right)}
\nc{\be}{\begin{equation}}
\nc{\ee}{\end{equation}}
\nc{\bea}{\begin{eqnarray}}
\nc{\eea}{\end{eqnarray}}
\nc{\bra}[1]{\langle {#1}|}
\nc{\ket}[1]{|{#1}\rangle}
\nc{\chit}{\raisebox{0.25ex}{$\chi$}}
\nc{\Db}{\mbox{\boldmath $D$}}
\nc{\Hb}{\mbox{\boldmath $H$}}
\nc{\Hc}{{\cal H}}
\nc{\Rc}{{\cal R}}
\nc{\Lc}{{\cal L}}
\nc{\Vc}{{\cal V}}
\nc{\Ib}{\mbox{\boldmath $I$}}
\nc{\qb}{\bar{q}}
\def\vvdots{\mathinner{\mkern1mu\raise1pt\vbox{\kern7pt\hbox{.}}\mkern2mu
  \raise4pt\hbox{.}\mkern2mu\raise7pt\hbox{.}\mkern1mu}}
\nc{\gauss}[2]{\left[\!\!\begin{array}{c} {#1}\\ {#2} \end{array}\!\!\right]}
\nc{\sbin}[2]{\left\{\!\!\!\begin{array}{c} {#1}\\ {#2} 
\end{array}\!\!\!\right\}}
\nc{\sbinlr}[2]{\Big\langle\!\!\begin{array}{c} {#1}\\ {#2} 
\end{array}\!\!\Big\rangle}
\nc{\bino}[2]{\left(\!\!\begin{array}{c} {#1}\\ {#2} \end{array}\!\!\right)}
\def\half {\mbox{$\textstyle \frac{1}{2}$}}
\def\vec#1{\mbox {\boldmath $#1$}}

\definecolor{lightblue}{rgb}{.7,.7,1}
\definecolor{lightlightblue}{rgb}{.85,.85,1}
\definecolor{purplish}{rgb}{.92,.83,.98}
\definecolor{yellowish}{rgb}{.95,.96,.7}

\def\leftarc#1{\psarc[linecolor=blue,linewidth=1.5pt]#1{.5}{90}{270}}
\def\rightarc#1{\psarc[linecolor=blue,linewidth=1.5pt]#1{.5}{-90}{90}}
\def\loopa{
\psframe[linewidth=.25pt](0,0)(1,1)
\psarc[linewidth=1.5pt,linecolor=blue](1,0){.5}{90}{180}
\psarc[linewidth=1.5pt,linecolor=blue](0,1){.5}{-90}{0}
}
\def\loopb{
\psframe[linewidth=.25pt](0,0)(1,1)
\psarc[linewidth=1.5pt,linecolor=blue](0,0){.5}{0}{90}
\psarc[linewidth=1.5pt,linecolor=blue](1,1){.5}{180}{270}
}

\def\facegrid#1#2{
\psframe[fillstyle=solid,fillcolor=lightlightblue,linewidth=0pt]#1#2
\psgrid[gridlabels=0pt,subgriddiv=1]#1#2}
\def\facegridblue#1#2{
\psframe[fillstyle=solid,fillcolor=lightblue,linewidth=0pt]#1#2
\psgrid[gridlabels=0pt,subgriddiv=1]#1#2}
\def\conn#1#2{
\psframe[linewidth=1pt,fillstyle=solid,fillcolor=lightlightblue]#1#2}

\begin{document}

\topmargin -5mm
\oddsidemargin 5mm

\begin{titlepage}
\setcounter{page}{0}

\vspace{8mm}
\begin{center}
{\huge {\bf Solvable Critical Dense Polymers}}

\vspace{10mm}
{\Large Paul A. Pearce}\ \ and\ \ {\Large J{\o}rgen Rasmussen}\\[.3cm]
{\em Department of Mathematics and Statistics, University of Melbourne}\\
{\em Parkville, Victoria 3010, Australia}\\[.4cm]
p.pearce@ms.unimelb.edu.au, j.rasmussen@ms.unimelb.edu.au

\end{center}

\vspace{10mm}
\centerline{{\bf{Abstract}}}
\vskip.4cm
\noindent
A lattice model of critical dense polymers is solved exactly for finite strips.
The model is the first member of the principal series of the recently 
introduced
logarithmic minimal models. The key to the solution is a functional equation in
the form of an inversion identity satisfied by the commuting 
double-row transfer matrices.
This is established directly in the planar Temperley-Lieb algebra and 
holds independently
of the space of link states on which the transfer matrices act. 
Different sectors are obtained
by acting on link states with $s-1$ defects where $s=1,2,3,\ldots$ is 
an extended Kac label.
The bulk and boundary free energies and finite-size corrections are 
obtained from the
Euler-Maclaurin formula.
The eigenvalues of the transfer matrix are classified by the physical combinatorics 
of the
patterns of zeros in the complex spectral-parameter plane.
This yields a selection rule for the physically relevant solutions to 
the inversion identity
and explicit finitized characters for the associated quasi-rational 
representations.
In particular, in the scaling limit, we confirm the central charge 
$c=-2$ and conformal
weights $\Delta_{s}=\frac{(2-s)^2-1}{8}$ for $s=1,2,3,\ldots$.
We also discuss a diagrammatic implementation of fusion and show with examples
how indecomposable representations arise. We examine the structure of these
representations and present a conjecture for the general fusion rules 
within our framework.
\end{titlepage}
\newpage
\renewcommand{\thefootnote}{\arabic{footnote}}
\setcounter{footnote}{0}

\section{Introduction}

Familiar materials such as plastics, nylon, polyester and plexiglass 
are made from polymers. Polymers consist of very long chain molecules 
with a large number of repeating structural units called monomers. 
Polymers exist in low- or high-temperature phases which are 
characterised as either dense or dilute. Polymers are {\em dense\/} 
if they fill a finite (non-zero) fraction of the available volume in 
the thermodynamic limit. In this paper, we are concerned with 
statistical models of dense polymers in two dimensions, both on the 
lattice and in the continuum. From the viewpoint of lattice 
statistical mechanics, polymers are of interest as prototypical 
examples of systems involving (extended) non-local degrees of 
freedom. It might be expected that the non-local nature of these 
degrees of freedom has a profound effect on the associated conformal 
field theory (CFT) obtained in the continuum scaling limit. Indeed, 
we confirm that the associated CFT is in fact {\em logarithmic\/} at 
least in the sense that certain representations of the dilatation 
Virasoro generator $L_0$ are non-diagonalizable and exhibit Jordan 
cells.

A simple model~\cite{SAW} of linear polymers, including excluded 
volume effects, is based on self-avoiding walks and can be obtained 
from the $O(n)$ model in the limit $n\to 0$. Similarly, a 
model~\cite{Potts0} of branching polymers (spanning trees) is 
obtained from the $Q$-state Potts model, or more precisely the 
dichromatic polynomial, in the limit $Q\to 0$. Although these 
approaches~\cite{Saleur87,Duplantier,SaleurSUSY,ReSa01} have yielded much 
insight into polymers, including the central charge, conformal weights and conformal characters, 
they rely on the properties of analytic 
continuation in $n$ or $Q$.

In this paper, we propose a Yang-Baxter integrable model of critical 
dense polymers on the square lattice whose properties are directly 
accessible {\em without\/} the need for analytic continuation. In 
fact, this model is the first member ${\cal LM}(1,2)$ of the 
principal series of the recently introduced integrable family of 
logarithmic minimal models~\cite{PRZ}. We study the model on a 
finite-width strip subject to certain boundary conditions. 
Remarkably, as for the square lattice Ising model, the transfer 
matrices of this model satisfy a single functional equation in the 
form of an inversion identity which is independent of the boundary 
conditions. This enables us to obtain the exact properties of the 
model on a finite lattice. The conformal properties are then readily 
accessible from finite-size corrections. In particular, in the 
continuum scaling limit, we confirm the central charge $c=-2$ and 
conformal
weights $\Delta_{s}=\frac{(2-s)^2-1}{8}$ where the extended Kac label 
$s=1,2,3,\ldots$ labels the boundary condition. Although all of these 
representations appear to be diagonalizable, we show how 
indecomposable representations are obtained by fusing boundary 
conditions.

In contrast to simple {\em rational\/} CFTs, in the case of {\em 
logarithmic\/} CFTs, there can be many different models with the same 
central charge and conformal weights. Specifically, the logarithmic 
theories at $c=-2$ include Hamiltonian walks on a Manhattan 
lattice~\cite{DuplantierDavid,Sedrakyan}, the rational triplet 
theory~\cite{GabK,FeiginEtAl,GabRunk}, symplectic 
fermions~\cite{Kau00}, the Abelian sandpile model~\cite{Ruelle}, 
dimers~\cite{dimers}, the travelling salesman problem~\cite{JRS} as 
well as branching polymers~\cite{branchpoly}. Although much is known 
about these models, it seems fair to say that at present the precise 
relation between these various models is still not clear.

The layout of this paper is as follows. In Section~2, we define our 
model of critical dense polymers and the planar Temperley-Lieb (TL) 
algebra~\cite{Jones} on which it is built. In Section~3, we set up 
double-row transfer matrices and prove the inversion identity 
directly in the planar TL algebra. We define vector spaces of link 
states on which the transfer matrices act and relate these to the 
boundary conditions. We solve the inversion identity to obtain the 
general solution for the eigenvalues of the transfer matrices. The 
physically relevant solutions, in the sectors with $\ell=s-1$ 
defects, are obtained empirically by selection rules encoded in 
suitably defined double-column diagrams. In Section~4, the 
finite-size corrections to the eigenvalues are obtained by applying 
the Euler-Maclaurin formula yielding the bulk and boundary free 
energies and confirming the central charge $c=-2$ and conformal
weights $\Delta_{s}=\frac{(2-s)^2-1}{8}$. In Section~5, the finitized 
conformal characters for $s=1,2,3,\ldots$ are obtained explicitly 
from finite-width partition functions on the strip. It is also 
verified that these yield the quasi-rational characters of~\cite{PRZ} 
in the continuum scaling limit. We discuss the Hamiltonian limit of 
the double-row transfer matrix in Section~6. In Section~7, 
we consider indecomposable representations. 
While it is well known~\cite{MartinBook} that the TL algebra admits indecomposable representations 
at roots of unity, this abstract result gives no hint as to how to construct 
these representations on the lattice or to relate them to physical models with physical boundary conditions. 
Here we exhibit the indecomposable representations of the dilatation 
generator $L_0$ arising from the lattice Hamiltonian and discuss how 
the fusion of such representations is implemented in terms of 
boundary conditions. 
The lattice approach to fusion is well established~\cite{LatticeFusion} in the context of boundary 
{\em rational} CFT and we extend the same principles here to {\em logarithmic} CFT. 
We conclude with some remarks and directions for 
future research in Section~8.

\section{Lattice Model}

\subsection{Critical dense polymers}

To model critical dense polymers on a finite strip, consider a square 
lattice consisting of $N$ columns and $N'$ rows of faces. An 
elementary face of the lattice can assume one of two configurations 
with different statistical weights
\be
\begin{pspicture}[.45](-.25,-.25)(1.25,1.25)
\facegrid{(0,0)}{(1,1)}
\put(0,0){\loopb}
\end{pspicture}
\quad\mbox{or}\quad
\begin{pspicture}[.45](-.25,-.25)(1.25,1.25)
\facegrid{(0,0)}{(1,1)}
\put(0,0){\loopa}
\end{pspicture}
\ee
where the arcs represent local segments of polymers. Since the 
polymer segments pass uniformly through each face, this is a model of 
{\em dense\/} polymers.
A typical configuration looks like
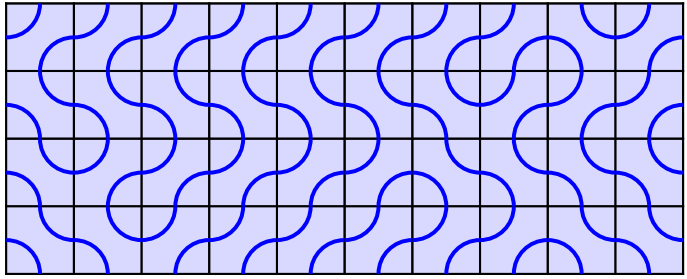
\psset{unit=.9cm}
\setlength{\unitlength}{.9cm}
\be
\begin{pspicture}[.45](0,-.3)(10,4.4)
\facegrid{(0,0)}{(10,4)}
\rput[bl](0,0){\loopb}
\rput[bl](1,0){\loopb}
\rput[bl](2,0){\loopa}
\rput[bl](3,0){\loopa}
\rput[bl](4,0){\loopa}
\rput[bl](5,0){\loopa}
\rput[bl](6,0){\loopa}
\rput[bl](7,0){\loopa}
\rput[bl](8,0){\loopb}
\rput[bl](9,0){\loopb}
\rput[bl](0,1){\loopb}
\rput[bl](1,1){\loopa}
\rput[bl](2,1){\loopa}
\rput[bl](3,1){\loopa}
\rput[bl](4,1){\loopa}
\rput[bl](5,1){\loopa}
\rput[bl](6,1){\loopb}
\rput[bl](7,1){\loopb}
\rput[bl](8,1){\loopb}
\rput[bl](9,1){\loopb}
\rput[bl](0,2){\loopb}
\rput[bl](1,2){\loopb}
\rput[bl](2,2){\loopb}
\rput[bl](3,2){\loopb}
\rput[bl](4,2){\loopb}
\rput[bl](5,2){\loopb}
\rput[bl](6,2){\loopb}
\rput[bl](7,2){\loopa}
\rput[bl](8,2){\loopa}
\rput[bl](9,2){\loopa}
\rput[bl](0,3){\loopa}
\rput[bl](1,3){\loopa}
\rput[bl](2,3){\loopa}
\rput[bl](3,3){\loopa}
\rput[bl](4,3){\loopa}
\rput[bl](5,3){\loopa}
\rput[bl](6,3){\loopa}
\rput[bl](7,3){\loopa}
\rput[bl](8,3){\loopb}
\rput[bl](9,3){\loopa}
\end{pspicture}
\label{typconf}
\ee
Once the boundary conditions have been specified, one is left with 
non-local degrees of freedom corresponding to a number of polymers. 
It is often convenient to think of these
degrees of freedom as non-local connectivities. Since polymers are 
not allowed to close on themselves, the fugacity $\beta$ assigned to 
loops in the associated loop gas vanishes.

Critical dense polymers, as defined above, corresponds to the first 
member ${\cal LM}(1,2)$ of the infinite principal series ${\cal 
LM}(m,m+1)$ of logarithmic minimal models introduced in~\cite{PRZ}.
Each logarithmic model is characterized by a crossing parameter
$\la$ which is a rational multiple of $\pi$ related to the loop 
fugacity $\beta$ by
\be
  \beta\ =\ 2\cos\la
\label{beta}
\ee
It was argued in~\cite{PRZ} that the scaling limits of these integrable lattice
models yield logarithmic CFTs. In the case of critical dense polymers,
\be
  \la\ =\ \frac{\pi}{2}
\label{la}
\ee
implying that loops appear with weight zero which is equivalent  to 
simply disallowing loops.

In this paper, we solve exactly our model of critical dense polymers 
on a strip. Remarkably, we can solve our model
for arbitrary finite sizes. The only other known case where this is 
possible is the Ising model on the square lattice~\cite{BaxBook,OPW}. 
The key to our solution is a functional equation in the form of an 
inversion identity reminiscent of the Ising model. This inversion 
identity is established directly in the planar Temperley-Lieb (TL) 
algebra~\cite{Jones}. As will be explained in Section~3, it therefore 
holds {\em independently} of the set of link states on which it acts, 
that is, for {\em all\/} matrix representations of the double-row 
transfer matrix.

\subsection{Planar Temperley-Lieb algebra}

To fix notation, we review here the basics of the planar TL algebra. 
We refer to \cite{Jones,PRZ} and references therein for more details.

An elementary face of the square lattice is assigned a face weight 
according to the configuration of the face.  The weights for the two 
possible configurations can be combined into a single face weight as
\psset{unit=.9cm}
\setlength{\unitlength}{.9cm}
\be
\begin{pspicture}[.45](-.5,-.1)(1.25,1.1)
\facegrid{(0,0)}{(1,1)}
\psarc[linewidth=.5pt](0,0){.15}{0}{90}
\rput(.5,.5){\small $u$}
\end{pspicture}
=\ \sin u\!\!
\begin{pspicture}[.45](-.5,-.1)(1.25,1.1)
\facegrid{(0,0)}{(1,1)}
\put(0,0){\loopb}
\end{pspicture}
+\ \cos u\!\!
\begin{pspicture}[.45](-.5,-.1)(1.25,1.1)
\facegrid{(0,0)}{(1,1)}
\put(0,0){\loopa}
\end{pspicture}
\label{u}
\ee
where the lower left corner has been marked to fix the orientation of 
the square.

For the purposes of this paper, the planar TL algebra is a 
diagrammatic algebra built up from elementary faces.
The faces are connected such that the midpoint of an outer edge of a 
face, called a node,
can be linked to a node of
any other (or even the same)
face as long as the total set of links make up a {\em 
non-intersecting planar} web of connections.
Two basic local properties of the planar TL algebra are the inversion relation
\psset{unit=.9cm}
\setlength{\unitlength}{.9cm}
\be
\begin{pspicture}[.45](-.5,0.75)(4,3.25)
\pspolygon[fillstyle=solid,fillcolor=lightlightblue](0,2)(1,1)(2,2)(1,3)(0,2)
\pspolygon[fillstyle=solid,fillcolor=lightlightblue](2,2)(3,1)(4,2)(3,3)(2,2)
\psarc[linewidth=.5pt](0,2){.15}{-45}{45}
\psarc[linewidth=.5pt](2,2){.15}{-45}{45}
\psarc[linecolor=blue,linewidth=1.5pt](2,2){.7}{45}{135}
\psarc[linecolor=blue,linewidth=1.5pt](2,2){.7}{-135}{-45}
\rput(1,2){\small $v$}
\rput(3,2){\small $\!-v$}
\end{pspicture}
  \ \ \ =\ \ \cos^2\!v\ \
\begin{pspicture}[.45](1,0.75)(4,3.25)
\pspolygon[fillstyle=solid,fillcolor=lightlightblue](1,2)(2,1)(3,2)(2,3)(1,2)
\psarc[linecolor=blue,linewidth=1.5pt](2,1){.7}{45}{135}
\psarc[linecolor=blue,linewidth=1.5pt](2,3){.7}{-135}{-45}
\end{pspicture}
\label{Inv}
\ee
and the Yang-Baxter equation (YBE)~\cite{BaxBook}
\psset{unit=.9cm}
\setlength{\unitlength}{.9cm}
\be
\begin{pspicture}[.45](-.5,0.75)(4,3.25)
\facegrid{(2,1)}{(3,3)}
\pspolygon[fillstyle=solid,fillcolor=lightlightblue](0,2)(1,1)(2,2)(1,3)(0,2)
\psarc[linewidth=.5pt](0,2){.15}{-45}{45}
\psline[linecolor=blue,linewidth=1.5pt](1.5,1.5)(2,1.5)
\psline[linecolor=blue,linewidth=1.5pt](1.5,2.5)(2,2.5)
\psarc[linewidth=.5pt](2,1){.15}{0}{90}
\psarc[linewidth=.5pt](2,2){.15}{0}{90}
\rput(2.5,1.5){\small $u$}
\rput(2.5,2.5){\small $v$}
\rput(1,2){\small $u-v$}
\end{pspicture}
  \!\!\! =\
\begin{pspicture}[.45](-.5,0.75)(4,3.25)
\facegrid{(0,1)}{(1,3)}
\pspolygon[fillstyle=solid,fillcolor=lightlightblue](1,2)(2,1)(3,2)(2,3)(1,2)
\psarc[linewidth=.5pt](1,2){.15}{-45}{45}
\psline[linecolor=blue,linewidth=1.5pt](1,1.5)(1.5,1.5)
\psline[linecolor=blue,linewidth=1.5pt](1,2.5)(1.5,2.5)
\psarc[linewidth=.5pt](0,1){.15}{0}{90}
\psarc[linewidth=.5pt](0,2){.15}{0}{90}
\rput(0.5,1.5){\small $v$}
\rput(0.5,2.5){\small $u$}
\rput(2,2){\small $u-v$}
\end{pspicture}
\label{YB}
\ee
These are identities for 2- and 3-tangles, respectively, where a 
$k$-tangle is an
arrangement of faces with $2k$ free nodes. The identities are established
by writing out all the possible configurations, while keeping track of the
associated weights, and collecting them in classes according to
connectivities. The left side of (\ref{Inv}), for example, thus corresponds
to a sum of four terms of which one vanishes since $\beta=0$. The 
remaining three terms
fall into the two connectivity classes
\be
\begin{pspicture}[.45](-.5,0.75)(4,3.25)
\pspolygon[fillstyle=solid,fillcolor=lightlightblue](0,2)(1,1)(2,2)(1,3)(0,2)
\pspolygon[fillstyle=solid,fillcolor=lightlightblue](2,2)(3,1)(4,2)(3,3)(2,2)
\psarc[linecolor=blue,linewidth=1.5pt](1,3){.7}{-135}{-45}
\psarc[linecolor=blue,linewidth=1.5pt](2,2){.7}{-135}{135}
\psarc[linecolor=blue,linewidth=1.5pt](1,1){.7}{45}{135}
\psarc[linecolor=blue,linewidth=1.5pt](4,2){.7}{135}{225}
\end{pspicture}
\ = \!\!\!\!
\begin{pspicture}[.45](-.5,0.75)(4,3.25)
\pspolygon[fillstyle=solid,fillcolor=lightlightblue](0,2)(1,1)(2,2)(1,3)(0,2)
\pspolygon[fillstyle=solid,fillcolor=lightlightblue](2,2)(3,1)(4,2)(3,3)(2,2)
\psarc[linecolor=blue,linewidth=1.5pt](2,2){.7}{45}{315}
\psarc[linecolor=blue,linewidth=1.5pt](0,2){.7}{-45}{45}
\psarc[linecolor=blue,linewidth=1.5pt](3,1){.7}{45}{135}
\psarc[linecolor=blue,linewidth=1.5pt](3,3){.7}{-135}{-45}
\end{pspicture}
\quad\ \mbox{or}\ \ \!\!
\begin{pspicture}[.45](-.5,0.75)(4,3.25)
\pspolygon[fillstyle=solid,fillcolor=lightlightblue](0,2)(1,1)(2,2)(1,3)(0,2)
\pspolygon[fillstyle=solid,fillcolor=lightlightblue](2,2)(3,1)(4,2)(3,3)(2,2)
\psarc[linecolor=blue,linewidth=1.5pt](2,2){.7}{45}{135}
\psarc[linecolor=blue,linewidth=1.5pt](2,2){.7}{-135}{-45}
\psarc[linecolor=blue,linewidth=1.5pt](1,1){.7}{45}{135}
\psarc[linecolor=blue,linewidth=1.5pt](1,3){.7}{-135}{-45}
\psarc[linecolor=blue,linewidth=1.5pt](3,1){.7}{45}{135}
\psarc[linecolor=blue,linewidth=1.5pt](3,3){.7}{-135}{-45}
\end{pspicture}
\ee
The weights accompanying the two equivalent configurations cancel
since $\cos v\sin(-v)+\sin v\cos(-v)=0$, while the last diagram
comes with the weight $\cos v\cos(-v)$ thereby yielding the identity 
(\ref{Inv}).

\section{Solution on a Finite Strip}

\subsection{Double-row transfer matrix}

Having introduced the planar TL algebra, we define diagrammatically
the double-row $N$-tangle
\psset{unit=.9cm}
\setlength{\unitlength}{.9cm}
\be
  \Db(u)\ =\ \frac{1}{\sin2u} \ \
  \begin{pspicture}[.5](-.5,.75)(8.5,3)
\facegrid{(0,1)}{(8,3)}
\leftarc{(0,2)}
\rightarc{(8,2)}
\rput(0.5,1.5){\small $u$}
\rput(7.5,1.5){\small $u$}
\rput(0.5,2.5){\small $\frac{\pi}{2}\!\!-\!\!u$}
\rput(7.5,2.5){\small $\frac{\pi}{2}\!\!-\!\!u$}
\rput(2.5,1.5){\small $\dots$}
\rput(5.5,1.5){\small $\dots$}
\rput(2.5,2.5){\small $\dots$}
\rput(5.5,2.5){\small $\dots$}
\psarc[linewidth=.5pt](0,1){.15}{0}{90}
\psarc[linewidth=.5pt](0,2){.15}{0}{90}
\psarc[linewidth=.5pt](7,1){.15}{0}{90}
\psarc[linewidth=.5pt](7,2){.15}{0}{90}
\end{pspicture}
\label{D}
\ee
consisting of $N$ two-columns, but suppress the
dependence on $N$. As discussed below, $\Db(u)$
has a natural matrix representation when acting vertically from below
on a given set of so-called link states.
We thus refer to it as the double-row transfer {\em matrix}, even though
it is defined as a planar $N$-tangle without reference to any matrix 
representation.
The normalization in (\ref{D}) ensures that
\be
  \lim_{u\rightarrow0}\Db(u)\ =\ \Ib\ =\
\begin{pspicture}[.5](-.25,.75)(8,3)
\conn{(0,1)}{(8,3)}
\multirput(.5,0)(1,0){8}{\psline[linecolor=blue,linewidth=1.5pt](0,1)(0,3)}
\end{pspicture}
\label{limD}
\ee
where $\Ib$ is the (vertical) identity operator linking every node on the upper
horizontal edge to the node directly below it on the lower horizontal edge.
Aside from the normalization, the double-row transfer matrix $\Db(u)$
is equivalent to the double-row transfer matrices introduced in \cite{PRZ}
for general fugacity $\beta$. Using diagrammatic arguments based on 
\cite{BPO}, it follows that $\Db(u)$ is crossing symmetric
\be
   \Db(\frac{\pi}{2}-u)\ =\ \Db(u)
\label{cross}
\ee
and that it gives rise to a commuting family of transfer matrices
where $[\Db(u),\Db(v)]=0$. Here the implied multiplication means 
vertical concatenation of the two $N$-tangles in the planar TL 
algebra.

\subsection{Inversion identity}

Remarkably, the double-row transfer matrix (\ref{D}) satisfies an inversion
identity in the planar TL algebra. This simple inversion identity is 
unique to critical dense polymers among the logarithmic minimal 
models~\cite{PRZ}.
\\[.3cm]
\noindent {\bf Inversion identity}\ \
{\em The planar} $N$-{\em tangle} $\Db(u)$ {\em defined in} (\ref{D}) 
{\em satisfies the inversion
identity}
\be
  \Db(u)\Db(u+\frac{\pi}{2})\ =\ 
\left(\frac{\cos^{2N}\!u-\sin^{2N}\!u}{\cos^2\!u-\sin^2\!u}
   \right)^{\!2}\Ib
\label{DDI}
\ee
{\em where} $\Ib$ {\em is the identity $N$-tangle} (\ref{limD}).
\\[.2cm]
{\bf Proof of inversion identity}\ \
The left side of (\ref{DDI}) corresponds diagrammatically to
\psset{unit=.75cm}
\setlength{\unitlength}{.75cm}
\bea
\vec D(u)\vec D(u+\frac{\pi}{2})&=&-\frac{1}{\sin^2 2u}\
\begin{pspicture}[.475](-.5,-.25)(8.5,4.25)
\facegrid{(0,0)}{(8,4)}
\leftarc{(0,1)}
\leftarc{(0,3)}
\rightarc{(8,1)}
\rightarc{(8,3)}
\rput(3.5,.5){\small $u$}
\rput(3.5,1.5){\small $\frac{\pi}{2}\!\!-\!\!u$}
\rput(3.5,2.5){\small $\frac{\pi}{2}\!\!+\!\!u$}
\rput(3.5,3.5){\small $-u$}
\end{pspicture}
\label{DDD}
\eea
where all faces have the standard orientation, cf. (\ref{u}) and (\ref{D}).
Employing the local inversion relation (\ref{Inv}) once followed by repeated
applications of the push-through property of the YBE (\ref{YB}),
transforms the right side of (\ref{DDD}) to
\bea
&&\begin{pspicture}[.475](-.5,-.25)(3.5,4.25)
\facegrid{(0,0)}{(3,4)}
\leftarc{(0,1)}
\leftarc{(0,3)}
\rightarc{(3,1)}
\rightarc{(3,3)}
\rput(1.5,.5){\small $u$}
\rput(1.5,1.5){\small $\frac{\pi}{2}\!\!-\!\!u$}
\rput(1.5,2.5){\small $\frac{\pi}{2}\!\!+\!\!u$}
\rput(1.5,3.5){\small $-u$}
\end{pspicture}
\;=\;\frac{1}{\cos^2 2u}\
\begin{pspicture}[.475](-.5,-.25)(10.5,4.25)
\facegrid{(0,0)}{(3,4)}
\facegrid{(7,0)}{(10,4)}
\psline[linecolor=blue,linewidth=1.5pt](3,.5)(7,.5)
\psline[linecolor=blue,linewidth=1.5pt](3,3.5)(7,3.5)
\psline[linecolor=blue,linewidth=1.5pt](3,1.5)(7,1.5)
\psline[linecolor=blue,linewidth=1.5pt](3,2.5)(7,2.5)
\pspolygon[fillstyle=solid,fillcolor=lightlightblue](3,2)(4,1)(5,2)(4,3)(3,2)
\pspolygon[fillstyle=solid,fillcolor=lightlightblue](5,2)(6,1)(7,2)(6,3)(5,2)
\psarc[linewidth=.5pt](3,2){.15}{-45}{45}
\psarc[linewidth=.5pt](5,2){.15}{-45}{45}
\leftarc{(0,1)}
\leftarc{(0,3)}
\rightarc{(10,1)}
\rightarc{(10,3)}
\rput(1.5,.5){\small $u$}
\rput(1.5,1.5){\small $\frac{\pi}{2}\!\!-\!\!u$}
\rput(1.5,2.5){\small $\frac{\pi}{2}\!\!+\!\!u$}
\rput(1.5,3.5){\small $-u$}
\rput(8.5,.5){\small $u$}
\rput(8.5,1.5){\small $\frac{\pi}{2}\!\!-\!\!u$}
\rput(8.5,2.5){\small $\frac{\pi}{2}\!\!+\!\!u$}
\rput(8.5,3.5){\small $-u$}
\rput(4,2){\small $2u$}
\rput(6,2){\small $\!-2u$}
\end{pspicture}\nonumber\\
&=&\!\frac{1}{\cos^2 2u}
\begin{pspicture}[.475](-.5,-.25)(10.5,4.25)
\psframe[linecolor=blue,framearc=1,linewidth=1.5pt](0,.5)(10,1.5)
\psframe[linecolor=blue,framearc=1,linewidth=1.5pt](0,2.5)(10,3.5)
\facegrid{(2,0)}{(8,4)}
\pspolygon[fillstyle=solid,fillcolor=lightlightblue](0,2)(1,1)(2,2)(1,3)(0,2)
\pspolygon[fillstyle=solid,fillcolor=lightlightblue](8,2)(9,1)(10,2)(9,3)(8,2)
\psarc[linewidth=.5pt](0,2){.15}{-45}{45}
\psarc[linewidth=.5pt](8,2){.15}{-45}{45}
\rput(4.5,.5){\small $u$}
\rput(4.5,1.5){\small $\frac{\pi}{2}\!\!+\!\!u$}
\rput(4.5,2.5){\small $\frac{\pi}{2}\!\!-\!\!u$}
\rput(4.5,3.5){\small $-u$}
\rput(1,2){\small $2u$}
\rput(9,2){\small $\!-2u$}
\end{pspicture}\nonumber\\
&=&\!\frac{1}{\cos^2 2u}\left\{-\sin^2 2u
\begin{pspicture}[.475](-.5,-.25)(5.5,4.25)
\psellipse[linecolor=blue,linewidth=1.5pt](1,2)(1.2,1.53)
\psellipse[linecolor=blue,linewidth=1.5pt](4,2)(1.2,1.53)
\leftarc{(1,2)}
\rightarc{(4,2)}
\facegrid{(1,0)}{(4,4)}
\rput(2.5,.5){\small $u$}
\rput(2.5,1.5){\small $\frac{\pi}{2}\!\!+\!\!u$}
\rput(2.5,2.5){\small $\frac{\pi}{2}\!\!-\!\!u$}
\rput(2.5,3.5){\small $-u$}
\end{pspicture}
\!+\sin 2u\cos 2u
\begin{pspicture}[.475](-.5,-.25)(5,4.25)
\psellipse[linecolor=blue,linewidth=1.5pt](1,2)(1.2,1.53)
\leftarc{(1,2)}
\rightarc{(4,1)}
\rightarc{(4,3)}
\facegrid{(1,0)}{(4,4)}
\rput(2.5,.5){\small $u$}
\rput(2.5,1.5){\small $\frac{\pi}{2}\!\!+\!\!u$}
\rput(2.5,2.5){\small $\frac{\pi}{2}\!\!-\!\!u$}
\rput(2.5,3.5){\small $-u$}
\end{pspicture}
\right.\nonumber\\
&&\mbox{}-\cos 2u\sin 2u\left.
\begin{pspicture}[.475](0,-.25)(5.5,4.25)
\psellipse[linecolor=blue,linewidth=1.5pt](4,2)(1.2,1.53)
\leftarc{(1,1)}
\leftarc{(1,3)}
\rightarc{(4,2)}
\facegrid{(1,0)}{(4,4)}
\rput(2.5,.5){\small $u$}
\rput(2.5,1.5){\small $\frac{\pi}{2}\!\!+\!\!u$}
\rput(2.5,2.5){\small $\frac{\pi}{2}\!\!-\!\!u$}
\rput(2.5,3.5){\small $-u$}
\end{pspicture}
+\cos^2 2u
\begin{pspicture}[.475](0,-.25)(5.5,4.25)
\leftarc{(1,1)}
\leftarc{(1,3)}
\rightarc{(4,1)}
\rightarc{(4,3)}
\facegrid{(1,0)}{(4,4)}
\rput(2.5,.5){\small $u$}
\rput(2.5,1.5){\small $\frac{\pi}{2}\!\!+\!\!u$}
\rput(2.5,2.5){\small $\frac{\pi}{2}\!\!-\!\!u$}
\rput(2.5,3.5){\small $-u$}
\end{pspicture}
\right\}
\label{DDDexp}
\eea
The last equality follows from an expansion in terms of connectivities
of the two faces with face weights $\pm2u$.
We now eliminate the last three diagrams by examining the 
consequences of having
a half-arc connecting the two left edges (or two right edges) of a two-column
appearing in the two lower or two upper rows. Expanding in terms of 
connectivities, we
find
\bea
\mbox{}\hspace{-.2in}
\begin{pspicture}[.475](-.75,-.25)(1,2.25)
\leftarc{(0,1)}
\facegrid{(0,0)}{(1,2)}
\rput(.5,1.5){\small $\frac{\pi}{2}\!\!+\!\!u$}
\rput(.5,.5){\small $u$}
\end{pspicture}
&=&\!\!-\sin u\cos u
\begin{pspicture}[.475](-.75,-.25)(1,2.25)
\leftarc{(0,1)}
\facegrid{(0,0)}{(1,2)}
\put(0,0){\loopa}
\put(0,1){\loopa}
\end{pspicture}
\;+\,\cos u\sin u
\begin{pspicture}[.475](-.75,-.25)(1,2.25)
\leftarc{(0,1)}
\facegrid{(0,0)}{(1,2)}
\put(0,0){\loopb}
\put(0,1){\loopb}
\end{pspicture}
\;-\,\sin^2 \!u
\begin{pspicture}[.475](-.75,-.25)(1,2.25)
\leftarc{(0,1)}
\facegrid{(0,0)}{(1,2)}
\put(0,0){\loopb}
\put(0,1){\loopa}
\end{pspicture}
\;+\,0\!\;
\begin{pspicture}[.475](-.75,-.25)(1,2.25)
\leftarc{(0,1)}
\facegrid{(0,0)}{(1,2)}
\put(0,0){\loopa}
\put(0,1){\loopb}
\end{pspicture}\nonumber\\
&=&\!\!-\sin^2 \!u
\begin{pspicture}[.475](-.75,-.25)(1,2.25)
\leftarc{(0,1)}
\facegrid{(0,0)}{(1,2)}
\put(0,0){\loopb}
\put(0,1){\loopa}
\end{pspicture}
\eea
and similarly
\bea
\begin{pspicture}[.475](0,-.25)(1.5,2.25)
\rightarc{(1,1)}
\facegrid{(0,0)}{(1,2)}
\rput(.5,1.5){\small $\frac{\pi}{2}\!\!+\!\!u$}
\rput(.5,.5){\small $u$}
\end{pspicture}\;=\,\cos^2 \!u
\begin{pspicture}[.475](-.25,-.25)(1.5,2.25)
\rightarc{(1,1)}
\facegrid{(0,0)}{(1,2)}
\put(0,0){\loopa}
\put(0,1){\loopb}
\end{pspicture}\;,\ \
\begin{pspicture}[.475](-.75,-.25)(1,2.25)
\leftarc{(0,1)}
\facegrid{(0,0)}{(1,2)}
\rput(.5,.5){\small $\frac{\pi}{2}\!\!-\!\!u$}
\rput(.5,1.5){\small $\!-u$}
\end{pspicture}\;=\,\cos^2 \!u
\begin{pspicture}[.475](-.75,-.25)(1,2.25)
\leftarc{(0,1)}
\facegrid{(0,0)}{(1,2)}
\put(0,0){\loopb}
\put(0,1){\loopa}
\end{pspicture}\;,\quad\ \
\begin{pspicture}[.5](0,-.25)(1.5,2.25)
\rightarc{(1,1)}
\facegrid{(0,0)}{(1,2)}
\rput(.5,.5){\small $\frac{\pi}{2}\!\!-\!\!u$}
\rput(.5,1.5){\small $\!-u$}
\end{pspicture}\;=\,-\sin^2 \!u
\begin{pspicture}[.475](-.25,-.25)(1.5,2.25)
\rightarc{(1,1)}
\facegrid{(0,0)}{(1,2)}
\put(0,0){\loopa}
\put(0,1){\loopb}
\end{pspicture}
\eea
We observe that such a half-arc will propagate. In all three diagrams 
under consideration (\ref{DDDexp}), this will eventually lead to a 
closed loop thus yielding a vanishing contribution.
Also due to the propagating property of the half-arcs,
the surviving diagram on the right side of (\ref{DDDexp}) may
be expanded as
\bea
&&\begin{pspicture}[.475](-1.25,-.25)(5.25,4.25)
\psellipse[linecolor=blue,linewidth=1.5pt](1,2)(1.2,1.53)
\psellipse[linecolor=blue,linewidth=1.5pt](4,2)(1.2,1.53)
\leftarc{(1,2)}
\rightarc{(4,2)}
\facegrid{(1,0)}{(4,4)}
\rput(2.5,.5){\small $u$}
\rput(2.5,1.5){\small $\frac{\pi}{2}\!\!+\!\!u$}
\rput(2.5,2.5){\small $\frac{\pi}{2}\!\!-\!\!u$}
\rput(2.5,3.5){\small $-u$}
\end{pspicture}
\;=\;\sin^{4N}\!u\;
\begin{pspicture}[.475](-1.25,-.25)(7.75,4.25)
\psellipse[linecolor=blue,linewidth=1.5pt](0,2)(1.2,1.53)
\psellipse[linecolor=blue,linewidth=1.5pt](6,2)(1.2,1.53)
\facegrid{(0,0)}{(6,4)}
\leftarc{(0,2)}
\rightarc{(6,2)}
\multiput(0,0)(1,0){6}{\loopb}
\multiput(0,1)(1,0){6}{\loopa}
\multiput(0,2)(1,0){6}{\loopa}
\multiput(0,3)(1,0){6}{\loopb}
\end{pspicture}\nonumber\\
&&\mbox{}+\cos^{4N}\!u\;
\begin{pspicture}[.475](-1.25,-.25)(7.75,4.25)
\psellipse[linecolor=blue,linewidth=1.5pt](0,2)(1.2,1.53)
\psellipse[linecolor=blue,linewidth=1.5pt](6,2)(1.2,1.53)
\facegrid{(0,0)}{(6,4)}
\leftarc{(0,2)}
\rightarc{(6,2)}
\multiput(0,0)(1,0){6}{\loopa}
\multiput(0,1)(1,0){6}{\loopb}
\multiput(0,2)(1,0){6}{\loopb}
\multiput(0,3)(1,0){6}{\loopa}
\end{pspicture}\nonumber\\
&&\mbox{}+
\begin{pspicture}[.475](-1.25,-.25)(10.25,4.25)
\psellipse[linecolor=blue,linewidth=1.5pt](0,2)(1.2,1.53)
\psellipse[linecolor=blue,linewidth=1.5pt](8,2)(1.2,1.53)
\facegrid{(0,0)}{(8,4)}
\leftarc{(0,2)}
\rightarc{(8,2)}
\multiput(1,2)(0,1){2}{\loopa}
\multiput(3,2)(0,1){2}{\loopb}
\multiput(4,0)(0,1){2}{\loopa}
\multiput(6,0)(0,1){2}{\loopb}
\rput(2.5,3.2){or}
\rput(5.5,1.2){or}
\psframe[linewidth=.9pt](1,2)(2,4)
\psframe[linewidth=.9pt](3,2)(4,4)
\psframe[linewidth=.9pt](4,0)(5,2)
\psframe[linewidth=.9pt](6,0)(7,2)
\end{pspicture}
\label{upordown}
\eea
There are two possible connectivities in the last
diagram for each two-column configuration in the lower or upper half, 
respectively.
Since the two first diagrams in (\ref{upordown}) both
represent the identity (\ref{limD}), we now focus on the last diagram.
Any given four-column configuration in this diagram can have only one of four
possible connectivities.
Reading from the left, we find by expanding in terms of these connectivities
that
\bea
\begin{pspicture}[.475](-1.25,-.25)(2.25,4.25)
\psellipse[linecolor=blue,linewidth=1.5pt](0,2)(1.2,1.53)
\facegrid{(0,0)}{(2,4)}
\leftarc{(0,2)}
\multiput(0,0)(0,1){4}{\loopa}
\end{pspicture}\;=\;
\begin{pspicture}[.475](-1.25,-.25)(2.25,4.25)
\psellipse[linecolor=blue,linewidth=1.5pt](0,2)(1.2,1.53)
\facegrid{(0,0)}{(2,4)}
\leftarc{(0,2)}
\multiput(0,0)(0,1){4}{\loopb}
\end{pspicture}\;=\;0\;,\qquad
\begin{pspicture}[.475](-1.25,-.25)(1.25,4.25)
\psellipse[linecolor=blue,linewidth=1.5pt](0,2)(1.2,1.53)
\psframe[fillstyle=solid,fillcolor=white,linecolor=white,linewidth=0pt](.5,0)(1.5,4)
\facegrid{(0,0)}{(1,4)}
\leftarc{(0,2)}
\multiput(0,0)(0,1){2}{\loopa}
\multiput(0,2)(0,1){2}{\loopb}
\end{pspicture}\;=\;0
\label{upup0}
\eea
This implies that a pair of nested half-arcs, as in the last diagram 
of (\ref{upordown}),
will propagate towards the right. The pair will not, however, 
propagate all the way to the
right to form a pair of closed loops since the sums over possible 
connectivities of the second
four-columns were required to establish the vanishing of the left
diagrams in (\ref{upup0}).
Continuing the evaluation in (\ref{upordown}), we thus have
\bea
\begin{pspicture}[.475](-1.25,-.25)(7.25,4.25)
\psellipse[linecolor=blue,linewidth=1.5pt](1,2)(1.2,1.53)
\psellipse[linecolor=blue,linewidth=1.5pt](6,2)(1.2,1.53)
\leftarc{(1,2)}
\rightarc{(6,2)}
\facegrid{(1,0)}{(6,4)}
\rput(3.5,.5){\small $u$}
\rput(3.5,1.5){\small $\frac{\pi}{2}\!\!+\!\!u$}
\rput(3.5,2.5){\small $\frac{\pi}{2}\!\!-\!\!u$}
\rput(3.5,3.5){\small $-u$}
\end{pspicture}\!&=&\!(\cos^{4N} \!u+\sin^{4N} \!u)\;
\begin{pspicture}[.475](-.25,-.25)(6.75,4.25)
\conn{(0,0)}{(5,4)}
\multirput(.5,0)(1,0){5}{\psline[linecolor=blue,linewidth=1.5pt](0,0)(0,4)}
\end{pspicture}\nonumber\\
&&\hspace{-1.5in}\mbox{}+\;
(\cos u\sin u)^{2(N-1)}\;
\begin{pspicture}[.475](-1.25,-.25)(6.75,4.25)
\psellipse[linecolor=blue,linewidth=1.5pt](0,2)(1.2,1.53)
\psellipse[linecolor=blue,linewidth=1.5pt](5,2)(1.2,1.53)
\facegrid{(0,0)}{(5,4)}
\leftarc{(0,2)}
\rightarc{(5,2)}
\multiput(0,0)(1,0){4}{\loopb}
\multiput(0,1)(1,0){4}{\loopb}
\multiput(0,2)(1,0){4}{\loopa}
\multiput(0,3)(1,0){4}{\loopa}
\facegridblue{(4,0)}{(5,4)}
\end{pspicture}
\label{elephant}
\eea
where
\bea
\begin{pspicture}[.475](-.25,-.25)(1.25,4.25)
\facegridblue{(0,0)}{(1,4)}
\end{pspicture}\;=\;-\cos^2 \!u\sin^2 \!u\;
\begin{pspicture}[.475](-.25,-.25)(1.25,4.25)
\facegrid{(0,0)}{(1,4)}
\multiput(0,0)(0,1){4}{\loopa}
\end{pspicture}
\quad\mbox{or}\quad
\begin{pspicture}[.475](-.25,-.25)(1.25,4.25)
\facegridblue{(0,0)}{(1,4)}
\end{pspicture}\;=\;-\cos^2 \!u\sin^2 \!u
\begin{pspicture}[.475](-.25,-.25)(1.25,4.25)
\facegrid{(0,0)}{(1,4)}
\multiput(0,0)(0,1){4}{\loopb}
\end{pspicture}
\label{or}
\eea
Combining the weights following from the various steps in this analysis
readily produces (\ref{DDI}) thus completing the proof. $\Box$

It is noted that one could have read the last diagram in (\ref{upordown})
from the right instead which would have involved identities obtained from
(\ref{upup0}) by reflection with respect to a vertical line. In this case,
the four-column (\ref{or}) would have appeared as the {\em first} 
four-column in
the last diagram in (\ref{elephant}) while the conclusion (\ref{DDI}) 
remains the same.

\subsection{Link states}

A family, labelled by $\ell$,  of matrix representations of $\Db(u)$ 
is obtained by acting
with $\Db(u)$ from below on link states with precisely $\ell$ defects.
It is recalled that $\Db(u)$ has $N$ free nodes on the upper horizontal
edge. A link state specifies how these $N$ nodes are linked together.
A node that is not linked to another node gives rise
to a defect which may be viewed as a link to the point (above) at infinity.
The following represents a link state with 11 nodes and three defects
\psset{unit=.65cm}
\setlength{\unitlength}{.65cm}
\be
\begin{pspicture}(11,2)
\psarc[linecolor=Maroon,linewidth=1.5pt](1,0){.5}{0}{180}
\psline[linecolor=Maroon,linewidth=1.5pt](2.5,0)(2.5,1.5)
\psarc[linecolor=Maroon,linewidth=1.5pt](5,0){.5}{0}{180}
\psarc[linecolor=Maroon,linewidth=1.5pt](5,0){1.5}{0}{180}
\psarc[linecolor=Maroon,linewidth=1.5pt](8,0){.5}{0}{180}
\psline[linecolor=Maroon,linewidth=1.5pt](9.5,0)(9.5,1.5)
\psline[linecolor=Maroon,linewidth=1.5pt](10.5,0)(10.5,1.5)
\end{pspicture}
\label{link}
\ee
Let ${\cal L}_{N,\ell}$ denote the vector space of link states with 
precisely $\ell$ defects. Its dimension is
\be
  \dim({\cal L}_{N,\ell})\ =\ 
\bin{N}{\frac{N-\ell}{2}}-\bin{N}{\frac{N-\ell-2}{2}}
\label{numberdef}
\ee
If the action of the planar TL algebra is unrestricted, then defects 
can be annihilated in pairs
\psset{unit=.75cm}
\setlength{\unitlength}{.75cm}
\be
\begin{pspicture}(13,3.5)
\facegrid{(0,0)}{(8,2)}
\rput[bl](0,0){\loopb}
\rput[bl](1,0){\loopb}
\rput[bl](2,0){\loopa}
\rput[bl](3,0){\loopa}
\rput[bl](4,0){\loopa}
\rput[bl](5,0){\loopa}
\rput[bl](6,0){\loopa}
\rput[bl](7,0){\loopa}
\rput[bl](0,1){\loopa}
\rput[bl](1,1){\loopa}
\rput[bl](2,1){\loopa}
\rput[bl](3,1){\loopa}
\rput[bl](4,1){\loopa}
\rput[bl](5,1){\loopb}
\rput[bl](6,1){\loopb}
\rput[bl](7,1){\loopa}
\psarc[linecolor=blue,linewidth=1.5pt](0,1){.5}{90}{270}
\psarc[linecolor=blue,linewidth=1.5pt](8,1){.5}{-90}{90}
\psarc[linecolor=Maroon,linewidth=1.5pt](1,2){.5}{0}{180}
\psline[linecolor=Maroon,linewidth=1.5pt](2.5,2)(2.5,3.5)
\psarc[linecolor=Maroon,linewidth=1.5pt](5,2){.5}{0}{180}
\psarc[linecolor=Maroon,linewidth=1.5pt](5,2){1.5}{0}{180}
\psline[linecolor=Maroon,linewidth=1.5pt](7.5,2)(7.5,3.5)
\rput(11,2.2){initial state:}
\rput(11,0.2){resulting state:}
\end{pspicture}
\ \ \
\begin{pspicture}(4.5,3.5)
\psarc[linecolor=Maroon,linewidth=1.5pt](.5,2){.25}{0}{180}
\psline[linecolor=Maroon,linewidth=1.5pt](1.25,2)(1.25,2.75)
\psarc[linecolor=Maroon,linewidth=1.5pt](2.5,2){.25}{0}{180}
\psarc[linecolor=Maroon,linewidth=1.5pt](2.5,2){.75}{0}{180}
\psline[linecolor=Maroon,linewidth=1.5pt](3.75,2)(3.75,2.75)
\psarc[linecolor=Maroon,linewidth=1.5pt](.5,0){.25}{0}{180}
\psarc[linecolor=Maroon,linewidth=1.5pt](2,0){.25}{0}{180}
\psarc[linecolor=Maroon,linewidth=1.5pt](2,0){.75}{0}{180}
\psarc[linecolor=Maroon,linewidth=1.5pt](3.5,0){.25}{0}{180}
\end{pspicture}
\ee

To associate a fixed number of defects $\ell$ with a boundary 
condition on the right (or left),
we can close the defects on the right (or left) and allow them to 
propagate down the edge of the strip.
In this case, the action of $\Db(u)$ is restricted to ${\cal 
L}_{N,\ell}$ by forbidding annihilations of any pair of defects.
The following illustrates this action of $\Db(u)$ on link states with 
$\ell=2$ defects for a particular
configuration with $N=8$
\vspace{-.5in}
\bea
\raisebox{-3\unitlength}{
\psset{unit=.75cm}
\setlength{\unitlength}{.75cm}
\begin{pspicture}(10,8.5)
\facegrid{(0,0)}{(8,4)}
\facegridblue{(8,0)}{(10,4)}
\psline[linecolor=blue,linestyle=dashed,dash=.25 
.25,linewidth=1.5pt](0,4)(10,4)
\psline[linecolor=blue,linewidth=1.5pt](8.5,0)(8.5,4)
\psline[linecolor=blue,linewidth=1.5pt](9.5,0)(9.5,4)
\psline[linecolor=blue,linewidth=1.5pt](8,.5)(8.4,.5)
\psline[linecolor=blue,linewidth=1.5pt](8.6,.5)(9.4,.5)
\psline[linecolor=blue,linewidth=1.5pt](9.6,.5)(10,.5)
\psline[linecolor=blue,linewidth=1.5pt](8,1.5)(8.4,1.5)
\psline[linecolor=blue,linewidth=1.5pt](8.6,1.5)(9.4,1.5)
\psline[linecolor=blue,linewidth=1.5pt](9.6,1.5)(10,1.5)
\psline[linecolor=blue,linewidth=1.5pt](8,2.5)(8.4,2.5)
\psline[linecolor=blue,linewidth=1.5pt](8.6,2.5)(9.4,2.5)
\psline[linecolor=blue,linewidth=1.5pt](9.6,2.5)(10,2.5)
\psline[linecolor=blue,linewidth=1.5pt](8,3.5)(8.4,3.5)
\psline[linecolor=blue,linewidth=1.5pt](8.6,3.5)(9.4,3.5)
\psline[linecolor=blue,linewidth=1.5pt](9.6,3.5)(10,3.5)
\rput[bl](0,0){\loopb}
\rput[bl](1,0){\loopb}
\rput[bl](2,0){\loopa}
\rput[bl](3,0){\loopa}
\rput[bl](4,0){\loopa}
\rput[bl](5,0){\loopa}
\rput[bl](6,0){\loopa}
\rput[bl](7,0){\loopa}
\rput[bl](0,1){\loopa}
\rput[bl](1,1){\loopa}
\rput[bl](2,1){\loopa}
\rput[bl](3,1){\loopa}
\rput[bl](4,1){\loopa}
\rput[bl](5,1){\loopa}
\rput[bl](6,1){\loopb}
\rput[bl](7,1){\loopb}
\rput[bl](0,2){\loopb}
\rput[bl](1,2){\loopb}
\rput[bl](2,2){\loopb}
\rput[bl](3,2){\loopb}
\rput[bl](4,2){\loopb}
\rput[bl](5,2){\loopb}
\rput[bl](6,2){\loopb}
\rput[bl](7,2){\loopa}
\rput[bl](0,3){\loopa}
\rput[bl](1,3){\loopa}
\rput[bl](2,3){\loopa}
\rput[bl](3,3){\loopa}
\rput[bl](4,3){\loopa}
\rput[bl](5,3){\loopa}
\rput[bl](6,3){\loopa}
\rput[bl](7,3){\loopa}
\psarc[linecolor=blue,linewidth=1.5pt](0,1){.5}{90}{270}
\psarc[linecolor=blue,linewidth=1.5pt](0,3){.5}{90}{270}
\psarc[linecolor=blue,linewidth=1.5pt](10,1){.5}{-90}{90}
\psarc[linecolor=blue,linewidth=1.5pt](10,3){.5}{-90}{90}
\psarc[linecolor=Maroon,linewidth=1.5pt](1,4){.5}{0}{180}
\psarc[linecolor=Maroon,linewidth=1.5pt](7,4){.5}{0}{180}
\psarc[linecolor=Maroon,linewidth=1.5pt](4,4){.5}{0}{180}
\psarc[linecolor=Maroon,linewidth=1.5pt](7,4){1.5}{0}{180}
\psarc[linecolor=Maroon,linewidth=1.5pt](6,3.105){3.62}{14}{166}
\end{pspicture}}
\label{r2}
\eea
We label this boundary condition by $s=\ell+1=1,2,3,\ldots$ and refer 
to this restricted action of $\Db(u)$ as a {\em block\/} or {\em 
sector\/} even though $\ell$ is not a quantum number.
In general, these blocks are not normal, not symmetric and may have 
repeated eigenvalues.
We have nevertheless observed and verified in many examples
that this restricted action of $\Db(u)$ is {\em diagonalizable}.
This is in contrast to the unrestricted action of $\Db(u)$ which in general is
{\em non-diagonalizable}. We will return to this issue when discussing
fusion in Section~7.

\subsection{Exact solution for the eigenvalues}

Once a matrix representation of $\Db(u)$ has been fixed,
the planar inversion identity (\ref{DDI}) for the specified representation
translates into a functional relation satisfied by the accompanying
eigenvalues $D(u)$. It reads
\be
  D(u)D(u+\frac{\pi}{2})\ =\ 
\left(\frac{\cos^{2N}\!u-\sin^{2N}\!u}{\cos^2\!u-\sin^2\!u}
   \right)^{\!2}
\label{LaLa}
\ee
where the normalization (\ref{limD}) and the crossing symmetry (\ref{cross})
yield the conditions
\be
  D(0)\ =\ 1,\qquad D(\frac{\pi}{2}-u)\ =\ D(u)
\label{La0}
\ee

The analysis of (\ref{LaLa}) depends on the parity of $N$ as we have 
the factorization
\be
  \frac{\cos^{2N}\!u-\sin^{2N}\!u}{\cos^2\!u-\sin^2\!u}\ =\ \begin{cases}
   \displaystyle\ \frac{N}{2^{N-1}}\prod_{j=1}^{\frac{N}{2}-1}
   \Big(\frac{1}{\sin^2\frac{j\pi}{N}}-\sin^2 2u\Big),&\mbox{$N$ even}\\[18pt]
  \displaystyle\ \frac{1}{2^{N-1}}\prod_{j=1}^{\frac{N-1}{2}}
   \Big(\frac{1}{\sin^2\frac{(2j-1)\pi}{2N}}-\sin^2 2u\Big),&\mbox{$N$ odd}
   \end{cases}
\label{cs}
\ee
It is now straightforward to solve the functional relation 
(\ref{LaLa}), subject to
(\ref{La0}), by sharing out the zeros. We find
\be
  D(u)\ = \
  \begin{cases}\displaystyle\ \frac{N}{2^{N-1}}\prod_{j=1}^{\frac{N}{2}-1}
   \Big(\frac{1}{\sin\frac{j\pi}{N}}+\eps_j\sin 2u\Big)
   \Big(\frac{1}{\sin\frac{j\pi}{N}}+\mu_j\sin 2u\Big),&\mbox{$N$ even}\\[18pt]
\displaystyle\ \frac{1}{2^{N-1}}\prod_{j=1}^{\frac{N-1}{2}}
   \Big(\frac{1}{\sin\frac{(2j-1)\pi}{2N}}+\eps_j\sin 2u\Big)
   \Big(\frac{1}{\sin\frac{(2j-1)\pi}{2N}}+\mu_j\sin 2u\Big),&\mbox{$N$ odd}
   \end{cases}
\label{N}
\ee
where $\eps_j^2=\mu_j^2=1$ for all $j$. The appearance of {\em two} sets of
parameters $\{\eps_j\}$ and $\{\mu_j\}$
stems from the overall squaring in (\ref{LaLa}).

For either parity of $N$ in (\ref{N}),
the maximum eigenvalue is obtained for $\eps_j=\mu_j=1$ for all $j$
and corresponds to the ground state in the associated sector.
Excited states are generated
by switching a (finite) number of the parameters $\eps_j,\mu_j$
from $1$ to $-1$.
The number of possible excitations following from
(\ref{N}) is $2^{N-2}$ and $2^{N-1}$, respectively,
clearly exceeding the number of link states with a
definite number of defects, cf.~(\ref{numberdef}).
We thus need a set of selection rules to determine which eigenvalues
actually appear in the spectrum and therefore are physical. This 
information is neatly encoded by specifying the patterns of complex 
zeros of $D(u)$ which are allowed.

Generally, the zeros of an eigenvalue $D(u)$ come in conjugate pairs 
in the complex $u$-plane
and appear with a periodicity $\pi$ in the real part of $u$. They follow from
(\ref{N}) and are collectively described by
\be
  u\ \in\ 
\big\{(2+\nu_j)\frac{\pi}{4}\pm\frac{i}{2}\ln\tan\frac{t_j}{2}\big\}\ 
+\ \pi\mathbb{Z}
\label{uzeros}
\ee
where $\nu_j$ is $\eps_j$ or $\mu_j$, and
where we have introduced the arguments
\be
  t_j\ =\ \left\{\begin{array}{lll} \frac{j\pi}{N},\ \ \ \ \ \ \ &N\ 
{\rm even}\\[8pt]
    \frac{(2j-1)\pi}{2N},\ \ \ \ \ \ \ &N\ {\rm odd}  \end{array} \right.
\label{tj}
\ee
A typical pattern for $N=12$ is
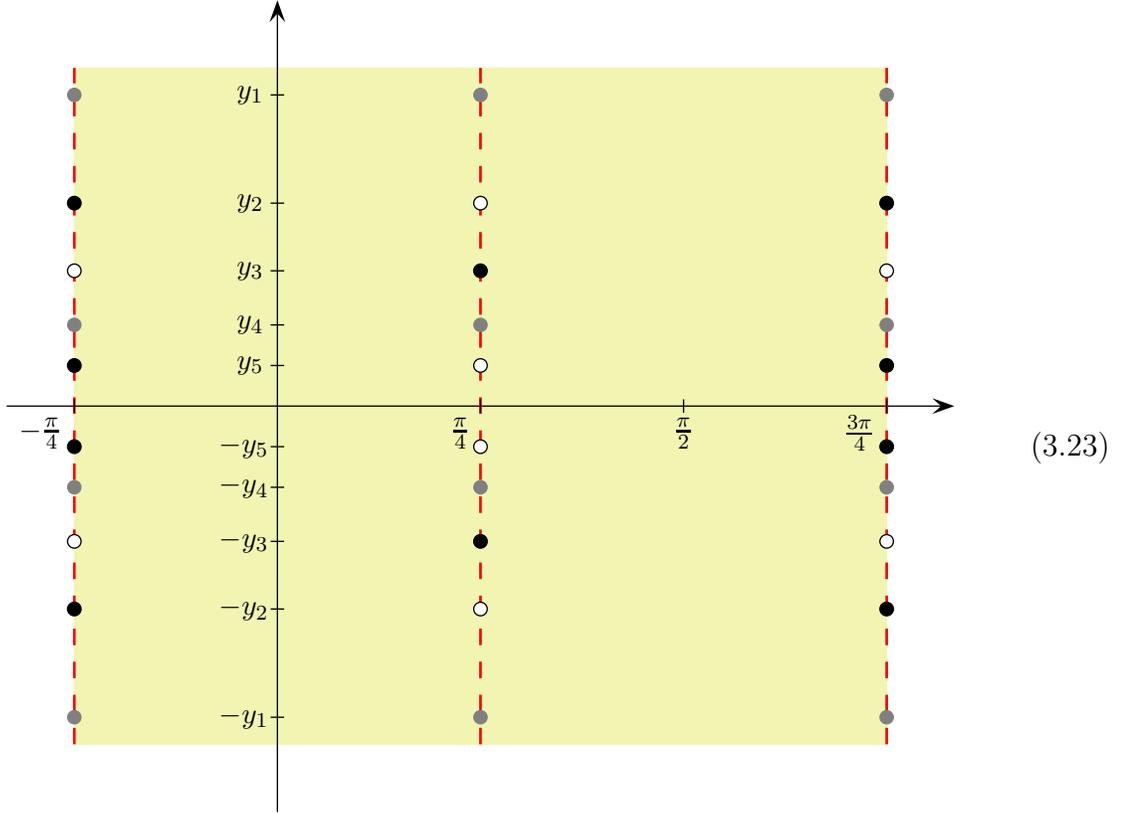
\psset{unit=.9cm}
\setlength{\unitlength}{.9cm}
\be
\begin{pspicture}[.45](-.25,-.25)(14,12)
\psframe[linecolor=yellowish,linewidth=0pt,fillstyle=solid,fillcolor=yellowish](1,1)(13,11)
\psline[linecolor=black,linewidth=.5pt,arrowsize=6pt]{->}(4,0)(4,12)
\psline[linecolor=black,linewidth=.5pt,arrowsize=6pt]{->}(0,6)(14,6)
\psline[linecolor=red,linewidth=1pt,linestyle=dashed,dash=.25 .25](1,1)(1,11)
\psline[linecolor=red,linewidth=1pt,linestyle=dashed,dash=.25 .25](7,1)(7,11)
\psline[linecolor=red,linewidth=1pt,linestyle=dashed,dash=.25 .25](13,1)(13,11)
\psline[linecolor=black,linewidth=.5pt](1,5.9)(1,6.1)
\psline[linecolor=black,linewidth=.5pt](7,5.9)(7,6.1)
\psline[linecolor=black,linewidth=.5pt](10,5.9)(10,6.1)
\psline[linecolor=black,linewidth=.5pt](13,5.9)(13,6.1)
\rput(.5,5.6){\small $-\frac{\pi}{4}$}
\rput(6.7,5.6){\small $\frac{\pi}{4}$}
\rput(10,5.6){\small $\frac{\pi}{2}$}
\rput(12.6,5.6){\small $\frac{3\pi}{4}$}
\psline[linecolor=black,linewidth=.5pt](3.9,6.6)(4.1,6.6)
\psline[linecolor=black,linewidth=.5pt](3.9,7.2)(4.1,7.2)
\psline[linecolor=black,linewidth=.5pt](3.9,8.0)(4.1,8.0)
\psline[linecolor=black,linewidth=.5pt](3.9,9.0)(4.1,9.0)
\psline[linecolor=black,linewidth=.5pt](3.9,10.6)(4.1,10.6)
\psline[linecolor=black,linewidth=.5pt](3.9,5.4)(4.1,5.4)
\psline[linecolor=black,linewidth=.5pt](3.9,4.8)(4.1,4.8)
\psline[linecolor=black,linewidth=.5pt](3.9,4.0)(4.1,4.0)
\psline[linecolor=black,linewidth=.5pt](3.9,3.0)(4.1,3.0)
\psline[linecolor=black,linewidth=.5pt](3.9,1.4)(4.1,1.4)
\rput(3.6,6.6){\small $y_5$}
\rput(3.6,7.2){\small $y_4$}
\rput(3.6,8.0){\small $y_3$}
\rput(3.6,9.0){\small $y_2$}
\rput(3.6,10.6){\small $y_1$}
\rput(3.5,5.4){\small $-y_5$}
\rput(3.5,4.8){\small $-y_4$}
\rput(3.5,4.0){\small $-y_3$}
\rput(3.5,3.0){\small $-y_2$}
\rput(3.5,1.4){\small $-y_1$}
\psarc[linecolor=black,linewidth=.5pt,fillstyle=solid,fillcolor=black](1,6.6){.1}{0}{360}
\psarc[linecolor=gray,linewidth=0pt,fillstyle=solid,fillcolor=gray](1,7.2){.1}{0}{360}
\psarc[linecolor=black,linewidth=.5pt,fillstyle=solid,fillcolor=white](1,8.0){.1}{0}{360}
\psarc[linecolor=black,linewidth=.5pt,fillstyle=solid,fillcolor=black](1,9.0){.1}{0}{360}
\psarc[linecolor=gray,linewidth=0pt,fillstyle=solid,fillcolor=gray](1,10.6){.1}{0}{360}
\psarc[linecolor=black,linewidth=.5pt,fillstyle=solid,fillcolor=white](7,6.6){.1}{0}{360}
\psarc[linecolor=gray,linewidth=0pt,fillstyle=solid,fillcolor=gray](7,7.2){.1}{0}{360}
\psarc[linecolor=black,linewidth=.5pt,fillstyle=solid,fillcolor=black](7,8.0){.1}{0}{360}
\psarc[linecolor=black,linewidth=.5pt,fillstyle=solid,fillcolor=white](7,9.0){.1}{0}{360}
\psarc[linecolor=gray,linewidth=0pt,fillstyle=solid,fillcolor=gray](7,10.6){.1}{0}{360}
\psarc[linecolor=black,linewidth=.5pt,fillstyle=solid,fillcolor=black](13,6.6){.1}{0}{360}
\psarc[linecolor=gray,linewidth=0pt,fillstyle=solid,fillcolor=gray](13,7.2){.1}{0}{360}
\psarc[linecolor=black,linewidth=.5pt,fillstyle=solid,fillcolor=white](13,8.0){.1}{0}{360}
\psarc[linecolor=black,linewidth=.5pt,fillstyle=solid,fillcolor=black](13,9.0){.1}{0}{360}
\psarc[linecolor=gray,linewidth=0pt,fillstyle=solid,fillcolor=gray](13,10.6){.1}{0}{360}
\psarc[linecolor=black,linewidth=.5pt,fillstyle=solid,fillcolor=black](1,5.4){.1}{0}{360}
\psarc[linecolor=gray,linewidth=0pt,fillstyle=solid,fillcolor=gray](1,4.8){.1}{0}{360}
\psarc[linecolor=black,linewidth=.5pt,fillstyle=solid,fillcolor=white](1,4.0){.1}{0}{360}
\psarc[linecolor=black,linewidth=.5pt,fillstyle=solid,fillcolor=black](1,3.0){.1}{0}{360}
\psarc[linecolor=gray,linewidth=0pt,fillstyle=solid,fillcolor=gray](1,1.4){.1}{0}{360}
\psarc[linecolor=black,linewidth=.5pt,fillstyle=solid,fillcolor=white](7,5.4){.1}{0}{360}
\psarc[linecolor=gray,linewidth=0pt,fillstyle=solid,fillcolor=gray](7,4.8){.1}{0}{360}
\psarc[linecolor=black,linewidth=.5pt,fillstyle=solid,fillcolor=black](7,4.0){.1}{0}{360}
\psarc[linecolor=black,linewidth=.5pt,fillstyle=solid,fillcolor=white](7,3.0){.1}{0}{360}
\psarc[linecolor=gray,linewidth=0pt,fillstyle=solid,fillcolor=gray](7,1.4){.1}{0}{360}
\psarc[linecolor=black,linewidth=.5pt,fillstyle=solid,fillcolor=black](13,5.4){.1}{0}{360}
\psarc[linecolor=gray,linewidth=0pt,fillstyle=solid,fillcolor=gray](13,4.8){.1}{0}{360}
\psarc[linecolor=black,linewidth=.5pt,fillstyle=solid,fillcolor=white](13,4.0){.1}{0}{360}
\psarc[linecolor=black,linewidth=.5pt,fillstyle=solid,fillcolor=black](13,3.0){.1}{0}{360}
\psarc[linecolor=gray,linewidth=0pt,fillstyle=solid,fillcolor=gray](13,1.4){.1}{0}{360}
\end{pspicture}
\label{uplane}
\ee
where
\be
  y_j\ =\ -\frac{i}{2}\ln\tan\frac{t_j}{2}
\ee
The infinite analyticity strip bounded by $u=-\frac{\pi}{4}$ and 
$u=\frac{3\pi}{4}$ is called the physical strip.
All zeros in the strip lie either on the boundary or on the vertical 
centre line (1-strings). In contrast to the usual situation in the 
context of Bethe ansatz, these zeros can occur either as single or 
double zeros.
A single zero is indicated by a grey dot while a double zero is indicated by
a black dot. Counting a double zero twice, the number of zeros with fixed
imaginary value $\pm y_j$ and real part either $\frac{\pi}{4}$ or 
$\frac{3\pi}{4}$
is two. This follows straightforwardly from (\ref{uzeros}) since 
$\nu_j$ can be either
$\epsilon_j$ or $\mu_j$, as depicted in (\ref{uplane}).
The fact that the zeros appear in complex conjugate pairs is now seen
to be a consequence of the crossing symmetry (\ref{La0}) and the 
periodicity in the $u$-plane.

It follows that the full pattern of zeros is encoded in the distribution
of 1-strings in the lower (or equivalently upper) half-plane.
This distribution is actually a sum of two, one governed by
excitations administered by $\epsilon$ and one by $\mu$.
The separation of the 1-strings
into contributions coming from $\epsilon$ and $\mu$, respectively,
is superfluous but nonetheless a helpful refinement later for the 
description of the
selection rules.
It is illustrated here
\psset{unit=.7cm}
\setlength{\unitlength}{.7cm}
\be
\begin{pspicture}[.45](-.25,-.25)(2,5)
\psframe[linewidth=0pt,fillstyle=solid,fillcolor=yellowish](0,0)(2,5)
\psarc[linecolor=black,linewidth=.5pt,fillstyle=solid,fillcolor=white](1,4.5){.1}{0}{360}
\psarc[linecolor=gray,linewidth=0pt,fillstyle=solid,fillcolor=gray](1,3.5){.1}{0}{360}
\psarc[linecolor=black,linewidth=.5pt,fillstyle=solid,fillcolor=black](1,2.5){.1}{0}{360}
\psarc[linecolor=black,linewidth=.5pt,fillstyle=solid,fillcolor=white](1,1.5){.1}{0}{360}
\psarc[linecolor=gray,linewidth=0pt,fillstyle=solid,fillcolor=gray](1,0.5){.1}{0}{360}
\end{pspicture}
\hspace{.6cm}\ \ \longleftrightarrow \hspace{.6cm}
\begin{pspicture}[.45](-.25,-.25)(2,5)
\psframe[linewidth=0pt,fillstyle=solid,fillcolor=yellowish](0,0)(2,5)
\psarc[linecolor=black,linewidth=.5pt,fillstyle=solid,fillcolor=white](0.5,4.5){.1}{0}{360}
\psarc[linecolor=black,linewidth=.5pt,fillstyle=solid,fillcolor=white](0.5,3.5){.1}{0}{360}
\psarc[linecolor=gray,linewidth=0pt,fillstyle=solid,fillcolor=gray](0.5,2.5){.1}{0}{360}
\psarc[linecolor=black,linewidth=.5pt,fillstyle=solid,fillcolor=white](0.5,1.5){.1}{0}{360}
\psarc[linecolor=black,linewidth=.5pt,fillstyle=solid,fillcolor=white](0.5,0.5){.1}{0}{360}
\psarc[linecolor=black,linewidth=.5pt,fillstyle=solid,fillcolor=white](1.5,4.5){.1}{0}{360}
\psarc[linecolor=gray,linewidth=0pt,fillstyle=solid,fillcolor=gray](1.5,3.5){.1}{0}{360}
\psarc[linecolor=gray,linewidth=0pt,fillstyle=solid,fillcolor=gray](1.5,2.5){.1}{0}{360}
\psarc[linecolor=black,linewidth=.5pt,fillstyle=solid,fillcolor=white](1.5,1.5){.1}{0}{360}
\psarc[linecolor=gray,linewidth=0pt,fillstyle=solid,fillcolor=gray](1.5,0.5){.1}{0}{360}
\end{pspicture}
\label{onetwo}
\ee
where the left side corresponds to (\ref{uplane}) while the two columns
to the right encode separately the $\epsilon$ and $\mu$ excitations.
Not all such double-column configurations will appear. A characterization of
a set of admissible two-column configurations provides a description 
of the selection rules.
This combinatorial designation of the physical states is termed 
`physical combinatorics'.
The details of the combinatorial content is the topic of our 
forthcoming paper \cite{PR_phys}.
A synopsis follows below and is used in our description of the selection rules
to be discussed subsequently. As we will see,
the double-column configurations also provide
a natural basis for defining the associated {\em finitized characters}.

\subsection{Double-column configurations}

A {\em single-column configuration} of height $M$ consists of $M$ 
sites arranged as
a column. The sites are labelled from the bottom by the
integers $1,\ldots,M$ and are weighted accordingly by $1,\ldots,M$.
A site can be occupied or unoccupied.
The {\em signature} $S$ of a single-column configuration
is constructed as the set of weights of the occupied sites in the configuration
listed in descending order. The {\em weight} $w$ of a configuration
is given by the sum of the signature entries
\be
  w(S)\ =\ \sum_jS_j
\label{wS}
\ee

A {\em double-column configuration} of height $M$ consists of a pair of
single-column configurations of height $M$. The notions of
signature and weight are readily
extended from the single-column case as illustrated by
the right side of (\ref{onetwo}). It has weight $w=11$
while the signature is
\be
  S\ =\ (L,R),\ \ \ \ \ \ \ L\ =\ (3),\ \ \ \ \ \ \ R\ =\ (4,3,1)
\label{SLSR}
\ee
Here $L$ and $R$ refer to the signatures of the left and right
single-column configurations, respectively.
With $m$ and $n$ indicating the number of occupied sites
in the left and right columns, respectively,
the weight $w$ of a double-column configuration is given by
\be
  w(L,R)\ =\ \sum_{j=1}^mL_j+\sum_{j=1}^nR_j
\label{wLR}
\ee

The set of single-column configurations of height $M$ admits
the partial ordering
\be
  S\ \preceq\ S'\ \ \ \ \ {\rm if}\ \ \ \ \ S_j\ \leq\ S'_j,\ \ j=1,\ldots,m
\label{order}
\ee
where $S$ and $S'$ are two such single-column configurations
with $m$ and $m'$ occupied sites, respectively.
A double-column configuration with signature $S=(L,R)$
is called {\em admissible} if $L\preceq R$.
This presupposes, in particular, that
\be
  0\ \leq\ m\ \leq\ n\ \leq\ M
\label{mnM}
\ee
The configuration (\ref{onetwo}) is seen to be admissible.
This convention for admissibility corresponds
to $\mu$ being dominant over $\epsilon$.
We let $A_{m,n}^M$ denote the set of admissible double-column configurations
with $m$ occupied sites in the left column
and $n$ occupied sites in the right column.
It is noted that the set $A_{m,n}^M$ is empty if and only if
one of the inequalities (\ref{mnM}) is violated.

There is a very simple `geometric' characterization of admissible
double-column configurations. As in the example
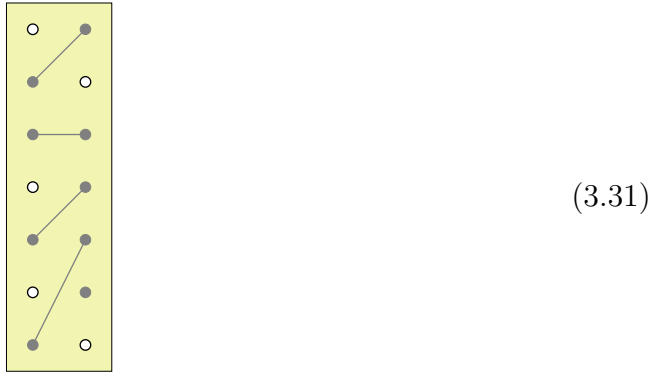
\psset{unit=.7cm}
\setlength{\unitlength}{.7cm}
\be
\begin{pspicture}[.45](-.25,-.25)(2,7.3)
\psframe[linewidth=0pt,fillstyle=solid,fillcolor=yellowish](0,0)(2,7)
\psarc[linecolor=black,linewidth=.5pt,fillstyle=solid,fillcolor=white](0.5,6.5){.1}{0}{360}
\psarc[linecolor=gray,linewidth=0pt,fillstyle=solid,fillcolor=gray](0.5,5.5){.1}{0}{360}
\psarc[linecolor=gray,linewidth=0pt,fillstyle=solid,fillcolor=gray](0.5,4.5){.1}{0}{360}
\psarc[linecolor=black,linewidth=.5pt,fillstyle=solid,fillcolor=white](0.5,3.5){.1}{0}{360}
\psarc[linecolor=gray,linewidth=0pt,fillstyle=solid,fillcolor=gray](0.5,2.5){.1}{0}{360}
\psarc[linecolor=black,linewidth=.5pt,fillstyle=solid,fillcolor=white](0.5,1.5){.1}{0}{360}
\psarc[linecolor=gray,linewidth=0pt,fillstyle=solid,fillcolor=gray](0.5,0.5){.1}{0}{360}
\psarc[linecolor=gray,linewidth=0pt,fillstyle=solid,fillcolor=gray](1.5,6.5){.1}{0}{360}
\psarc[linecolor=black,linewidth=.5pt,fillstyle=solid,fillcolor=white](1.5,5.5){.1}{0}{360}
\psarc[linecolor=gray,linewidth=0pt,fillstyle=solid,fillcolor=gray](1.5,4.5){.1}{0}{360}
\psarc[linecolor=gray,linewidth=0pt,fillstyle=solid,fillcolor=gray](1.5,3.5){.1}{0}{360}
\psarc[linecolor=gray,linewidth=0pt,fillstyle=solid,fillcolor=gray](1.5,2.5){.1}{0}{360}
\psarc[linecolor=gray,linewidth=0pt,fillstyle=solid,fillcolor=gray](1.5,1.5){.1}{0}{360}
\psarc[linecolor=black,linewidth=.5pt,fillstyle=solid,fillcolor=white](1.5,0.5){.1}{0}{360}
\psline[linecolor=gray,linewidth=.5pt](0.5,5.5)(1.5,6.5)
\psline[linecolor=gray,linewidth=.5pt](0.5,4.5)(1.5,4.5)
\psline[linecolor=gray,linewidth=.5pt](0.5,2.5)(1.5,3.5)
\psline[linecolor=gray,linewidth=.5pt](0.5,0.5)(1.5,2.5)
\end{pspicture}
\label{adm}
\ee
one draws line segments between the occupied sites of greatest weight
in the two columns, between the occupied sites of second-to-greatest
weight etc. Admissible double-column configurations are now characterized
by {\em not} involving line segments with a {\em negative slope}.
It could involve no line segments ($m=0$), while a line segment can only appear
with non-negative slope. The
example (\ref{adm}) is immediately recognized as being admissible.
As illustrated in (\ref{adm}), grey dots ultimately not linked to any 
other grey dot must
appear as the $n-m$ grey dots with lowest weight in the right column.

In preparation for the discussion of (finitized) characters below,
we associate a monomial $q^w$ to a double-column
configuration where $w$ is the weight of the configuration.
Here $q$ is a formal parameter but later it will be the modular parameter.
We define the polynomial $\sbinlr{M}{m,n}_{\!q}$
as the sum of the monomials associated to the elements of $A_{m,n}^M$.
In \cite{PR_phys}, we show that this polynomial admits the closed-form
expression
\be
  \sbinlr{M}{m,n}_{\!q}\ =\ q^{\hf m(m+1)+\hf n(n+1)}
   \bigg(\gauss{M}{m}_{\!q}\gauss{M}{n}_{\!q}-q^{n-m+1}\gauss{M}{n+1}_{\!q}
    \gauss{M}{m-1}_{\!q}\bigg)
\label{sbin}
\ee
where $\Big[\!\!\begin{array}{cc} k\\[-.18cm] j\end{array}\!\!\Big]_{\! q}$
is a $q$-binomial (Gaussian polynomial). These are generalized 
$q$-Narayana numbers~\cite{Narayana} to which they coincide when 
$m=n$.
Despite the minus sign, (\ref{sbin}) is actually {\em fermionic} in 
the sense that it is a
polynomial with only non-negative coefficients \cite{PR_phys}.

\subsection{Selection rules}

We now turn to the description of the selection rules.
They depend on the parity of $N$.
Based on a body of empirical data obtained
by examining the double-row
transfer matrices analytically as well as numerically,
we make the following conjecture.
\\[.3cm]
\noindent {\bf Selection rules}\ \
{\em For a system with $N$ columns and $\ell$ defects where} 
$N-\ell\equiv0$ mod $2$,
{\em the selection rules are equivalent to singling out the following sets
of admissible double-column configurations}
\be
  N\ {\rm even}:\ \ \ \bigcup_{m=0}^{\frac{N-\ell}{2}}
   \big(A_{m,m+\frac{\ell-2}{2}}^{\frac{N-2}{2}}\cup
      A_{m,m+\frac{\ell}{2}}^{\frac{N-2}{2}}\big),\ \ \ \ \ \ \
  N\ {\rm odd}:\ \ \ \bigcup_{m=0}^{\frac{N-\ell}{2}} 
A_{m,m+\frac{\ell-1}{2}}^{\frac{N-1}{2}}
\label{selruleA}
\ee
By construction, all these unions are disjoint unions.
It is noted that the case where $\ell=0$ is special since
$A_{m,m-1}^{\frac{N-2}{2}}=\emptyset$. As argued below, $\ell=0$ gives
rise to the identity field in the associated CFT.

\section{Finite-Size Corrections}

The partition function of critical dense polymers on a lattice of $N$ 
columns and $M$ double
rows is defined by
\be
  Z_{N,M}\ =\ \mathop{\rm Tr}\Db(u)^M\ =\ \sum_nD_{n}(u)^M
   \ =\ \sum_ne^{-ME_{n}(u)}
\label{ZNM}
\ee
where the sum is over all eigenvalues of $\Db(u)$ including possible
multiplicities and $E_{n}(u)$ is the energy associated to the eigenvalue
$D_{n}(u)$. A refinement of (\ref{ZNM}) to treat explicitly sectors 
with different numbers of defects is given in the next section.
Conformal invariance of the model in the continuum scaling limit would dictate
\cite{BCN,Aff} that the leading finite-size corrections for large $N$ 
are of the form
\be
  E_{n}(u)\ =\ -\ln D_{n}(u)\ \simeq\ 2Nf_{bulk}+f_{bdy}
   +\frac{2\pi\sin 2u}{N}\big(\!-\frac{c}{24}+\Delta+k\big)
\label{logLa}
\ee
Here $f_{bulk}$ is the bulk free energy per face, $f_{bdy}$ is the 
boundary or surface
free energy, while $c$ is the central charge of the CFT whose
spectrum of conformal weights is given by the possible values of $\D$
with excitations or descendants labelled by the non-negative integers $k$.
As the analysis below will confirm, the asymptotic behaviour of the
eigenvalues (\ref{N}) are in accordance with (\ref{logLa}).

The logarithms of the eigenvalues (\ref{N}) involves sums
of terms which are singular in the limit $N\rightarrow\infty$.
To remedy this, one can introduce the function
\be
  F(t)\ =\ \ln\big(\frac{t}{\sin t}+t\sin 2u\big)\ =\ \ln 
t+\ln\big(\frac{1}{\sin t}+\sin 2u\big)
\label{F}
\ee
which is well defined as $t\rightarrow 0$ and where the $u$ dependence
for simplicity has been formally suppressed.
This trick was employed in a similar analysis~\cite{OPW}
of the Ising model, and it allows us to examine the asymptotic
behaviour of the logarithms. Indeed, the Euler-Maclaurin formula now gives
\be
  \sum_{k=0}^mF(a+kh)\ \simeq\ \frac{1}{h}\int_a^bF(t)dt+\frac{1}{2}[F(b)+F(a)]
   +\frac{h}{12}[F'(b)-F'(a)]
\label{EM}
\ee
where $b=a+mh$, and due to the split (\ref{F}),
we also need to approximate the logarithm of the gamma function
\be
  \ln\Gamma(y)\ \simeq\ (y-\half)\ln y-y+\half\ln(2\pi)+\frac{1}{12y}
\label{Gamma}
\ee

As a partial evaluation of the energies (\ref{logLa}), we find
\be
  \ln\prod_j\Big(\frac{1}{\sin t_j}+\eps_j\sin 2u\Big)
  \ \simeq\ \sum_j\big(F(t_j)-\ln t_j\big)-2\sin 2u\sum_{j\in{\cal E}}t_j
\label{sumE}
\ee
where ${\cal E}$ is the subset of $j$ indices for which
$\eps_j=-1$ and ${\cal E}$ remains finite as $N\rightarrow\infty$.
Likewise, ${\cal M}$ denotes the subset of $j$ indices for which
$\mu_j=-1$ and it also remains finite as $N\rightarrow\infty$.
The argument $t_j$ is defined in (\ref{tj}).
We finally find
\bea
  E_{n}(u)&\simeq&2Nf_{bulk}+f_{bdy}\nn
  &+&\frac{2\pi\sin 2u}{N}
\begin{cases}
\displaystyle\
   \Big(\frac{2}{24}+\sum_{j\in{\cal E}_{N,n}}j
    +\sum_{j\in{\cal M}_{N,n}}j\Big),&\mbox{$N$ even}\\[18pt]
\displaystyle\
   \Big(\frac{2}{24}
    -\frac{1}{8}+\sum_{j\in{\cal E}_{N,n}}(j-\half)
    +\sum_{j\in{\cal M}_{N,n}}(j-\half)\Big),&\mbox{$N$ odd}
    \end{cases}
\label{logLafin}
\eea
In these expressions, the bulk free energy per face is given by
\be
  f_{bulk}\ =\ \ln\sqrt{2}-\frac{1}{\pi}\int_{0}^{\pi/2}
   \ln\Big(\frac{1}{\sin t}+\sin 2u\Big)dt
\label{fb}
\ee
whereas the boundary free energy is given by
\be
  f_{bdy}\ =\ \ln(1+\sin 2u)
\label{fs}
\ee

\section{Conformal Field Theory}

The next task and our main objective is to calculate analytically the 
conformal data
associated with our model of critical dense polymers.
By comparing the expressions  (\ref{logLafin}) with the finite-size 
corrections (\ref{logLa}),
the central charge is readily seen to be
\be
  c\ =\ -2
\label{c}
\ee
while the lowest conformal weights for given parity of $N$ are
$\D=0$ and $\D=-\frac{1}{8}$, respectively.
The full spectrum of conformal weights and their excitations are
discussed in the following.

\subsection{Finitized characters}

It is recalled that the matrix representation of the double-row
transfer matrix depends on which space $\Db(u)$ is acting.
Since the action of $\Db(u)$ on ${\cal L}_N$
is upper block-triangular, the partition
function (\ref{ZNM}) may be written
\be
  Z_{N,M}\ =\ \sum_{s}\sum_n D_{n}^{(s)}(u)^M
   \ =\ \sum_{s}\sum_ne^{-ME_{n}^{(s)}(u)}
\label{ZNMs}
\ee
where $D_{n}^{(s)}(u)$ and $E_{n}^{(s)}(u)$ are the eigenvalues
and associated energies obtained by restricting to the $\ell=s-1$ sector.
Likewise, information on the excitations or descendants are contained
in the appropriate sets ${\cal E}_{N,n}^{(s)}$ and ${\cal M}_{N,n}^{(s)}$,
cf.~(\ref{logLafin}).
Here we have introduced the extended Kac label $s$ related to the number of
defects $\ell$ by
\be
  s\ =\ \ell+1
\label{s}
\ee

Now, the link between the lattice model and the characters of the
CFT is governed by the modular parameter $q$ defined by
\be
  q\ =\ e^{-2\pi\tau},\ \ \ \ \ \ \ \tau\ =\ \frac{M}{N}\sin 2u
\label{tau}
\ee
The ratio $M/N$ is the aspect ratio and $\vartheta=2u$ is the 
anisotropy angle related to the geometry of the lattice.
As already discussed, the spectrum of the CFT is extracted
from the eigenvalues,
that is, the specification of ${\cal E}_{N,n}^{(s)}$ and ${\cal M}_{N,n}^{(s)}$
for all $s$. This amounts to determining appropriate selection rules.
The relevant rules are stated explicitly in (\ref{selruleA})
where $\ell=s-1$.

For finite $N$, we thus find that the full spectrum can be organized in
{\em finitized characters} $\chit_{s}^{(N)}(q)$
and we find that they are given by
\bea
  \chi_{s}^{(N)}(q)\ =\
  \begin{cases}\displaystyle\ q^{\frac{1}{12}}\sum_{m=0}^{\frac{N-s+1}{2}}
   \Big(\sbinlr{\frac{N-2}{2}}{m,m+\frac{s-3}{2}}_{\!q}
   +\sbinlr{\frac{N-2}{2}}{m,m+\frac{s-1}{2}}_{\!q}\Big),&\mbox{ $s$ 
odd}\\[18pt]
\displaystyle\  q^{-\frac{1}{24}-\frac{s-2}{4}}
   \sum_{m=0}^{\frac{N-s+1}{2}}
    \sbinlr{\frac{N-1}{2}}{m,m+\frac{s-2}{2}}_{\!q}q^{-m},&\mbox{$s$ even}
    \end{cases}
\label{ferm}
\eea
where some terms may vanish, cf.~(\ref{mnM}). It is emphasized that 
these are {\em fermionic} character expressions.

In \cite{PR_phys}, we prove that the finitized characters 
(\ref{ferm}) can be written collectively as
\bea
  \chi_{s}^{(N)}(q)&=&q^{-\frac{c}{24}+\Delta_{s}}
\Big(\gauss{N}{\frac{N-s+1}{2}}_{\!q}-q^{s}\gauss{N}{\frac{N-s-1}{2}}_{\!q}\Big)
\label{finchi}
\eea
where
\be
  \D_{s}\ =\ \frac{s^2-4s+3}{8},\ \ \ \ \ \ \ s\in\mathbb{Z}_>
\label{Ds}
\ee
This corresponds to the first column of the extended Kac table
\psset{unit=.7cm}
\setlength{\unitlength}{.7cm}
\be
\begin{pspicture}[.45](-.25,0)(1,7.25)
\psframe[fillstyle=solid,fillcolor=purplish,linewidth=0.5pt](0,0)(1,7)
\psgrid[gridlabels=0pt,subgriddiv=1]
\rput(.5,.5){$0$}
\rput(.5,1.5){$-\frac{1}{8}$}
\rput(.5,2.5){$0$}
\rput(.5,3.5){$\frac{3}{8}$}
\rput(.5,4.5){$1$}
\rput(.5,5.5){$\frac{15}{8}$}
\rput(.5,6.65){$\vdots$}
\end{pspicture}
\label{kac}
\ee
Setting, $\D_s=\D_{1,s}$, this agrees with the general ${\cal 
LM}(p,p')$ formula
\be
   \D_{r,s}\ =\ \frac{(rp'-sp)^2-(p'-p)^2}{4pp'}
\label{Drs}
\ee
with $(p,p')=(1,2)$ for critical dense polymers.
This explicitly confirms the form of the finitized characters in \cite{PRZ}
for the case $(p,p')=(1,2)$ with $r=1$.
Finitized characters like (\ref{finchi}) were introduced first 
in~\cite{Mel,Ber}.

By construction, a finitized character must contain information on
the dimensionality of the vector space of states. Here this is reflected in
the formula
\be
  \lim_{q\rightarrow1}\chi_{s}^{(N)}(q)=\mbox{dim}\,{\cal L}_{N,s-1}
\ee

\subsection{Spectrum of $c=-2$ CFT}

In the continuum scaling limit, the finitized characters carry over 
to the full characters of the associated CFT.
We have already identified the central charge $c=-2$ and the conformal weights
(\ref{Ds}) of this CFT while the associated characters are given by
\be
   \chit_{s}(q)\ =\ \lim_{N\to\infty}\chi_{s}^{(N)}(q)\ =\ 
\frac{q^{\frac{(s-2)^2}{8}}}{\eta(q)}
    (1-q^s)
\label{chis}
\ee
where the Dedekind eta function is defined by
\be
   \eta(q)\ =\ q^{\frac{1}{24}}\prod_{m=1}^\infty(1-q^m)
\label{eta}
\ee

The character $\chit_s(q)$ is recognized as the Virasoro character of 
a particular quasi-rational
highest-weight representation of the Virasoro algebra.
To appreciate this, we recall that
the Verma module $V_{r,s}$ of a Virasoro highest-weight 
representation of highest weight
(\ref{Drs})
is defined for all $r,s\in\mathbb{Z}_>$ and $p,p'$ two coprime 
positive integers.
The associated quotient module
\be
  Q_{r,s}\ =\ V_{r,s}/V_{r,-s}
\label{Q}
\ee
is also defined for every pair of positive integers $r,s$ and corresponds
to a quasi-rational representation.
The character of such a quasi-rational representation is given by
\be
  \chit_{r,s}(q)\ =\ \frac{q^{\frac{1-c}{24}}}{\eta(q)}\left(q^{\D_{r,s}}
     -q^{\D_{r,-s}}\right)
    \ =\ \frac{q^{\frac{1-c}{24}}}{\eta(q)}q^{\D_{r,s}}\left(1-q^{rs}\right)\nn
\label{chi}
\ee
where $c=1-6\frac{(p'-p)^2}{pp'}$.
Critical dense polymers correspond to $(p,p')=(1,2)$ in which case
$\chit_{1,s}(q)=\chit_s(q)$, as already indicated.

So far for critical dense polymers, from the lattice approach, we 
have only identified
sectors with $r=1$. The associated quasi-rational characters 
correspond to representations
$(1,s)$ which are only irreducible for $s=1$ or $s$ even.
{}From the point of view of CFT, one conventionally works with 
irreducible representations
as the natural building blocks. This is indeed the case in~\cite{GK} 
where the complete
set of irreducible representations necessarily involve sectors 
corresponding to $r>1$, even
after identification of irreducible representations with identical 
characters following the classification
of irreducible characters in the appendix of \cite{PRZ}.
A lattice analysis of the set of irreducible representations will 
appear elsewhere.

A variety of character identities exist for the set of quasi-rational 
representations. In particular,
every quasi-rational character (\ref{chi}) with $r>1$
can be written as a linear combination of characters
with $r=1$
\be
  \chit_{1+k,s}(q)\ =\ \sum_{j=0}^k\chit_{s+4j-2k}(q)
\label{r1}
\ee
where it is implicit that
\be
  \chit_{0}(q)\ =\ 0,\ \ \ \ \ \ \ \chit_{-s}(q)\ =\ -\chit_{s}(q)
\label{chi0}
\ee
This extends to general $r,s$ and coprime $p,p'$ \cite{PR_phys}
by admitting linear combinations of the $p$ left-most columns in the 
extended Kac table
\be
  \chit_{r+kp,s}(q)\ =\ \sum_{j=0}^k\chi_{r,s+(2j-k)p'}(q)
    +\sum_{j=0}^{k-1}\chi_{p-r,s+(2j+1-k)p'}(q)
\label{chir}
\ee
employing a straightforward extension of (\ref{chi0}).
It is emphasized, though, that these character identities are blind
to any Jordan cells, cf.~Section~7.

\section{Hamiltonian Limit}

The Hamiltonian limit of the double-row transfer matrix $\Db(u)$ is 
defined in the
planar algebra as the $N$-tangle appearing as the leading non-trivial 
term in an
expansion with respect to $u$, that is,
\be
  \Db(u)\ =\ \Ib\ -\ 2u\Hb\ +\ {\cal O}(u^2)
\label{DIH}
\ee
It follows that
\psset{unit=.6cm}
\setlength{\unitlength}{.6cm}
\be
  -\Hb\ =\
\begin{pspicture}[.45](-.25,.75)(6,3)
\conn{(0,1)}{(6,3)}
\psline[linecolor=blue,linewidth=1.5pt](2.5,1)(2.5,3)
\psline[linecolor=blue,linewidth=1.5pt](3.5,1)(3.5,3)
\psline[linecolor=blue,linewidth=1.5pt](5.5,1)(5.5,3)
\psarc[linecolor=blue,linewidth=1.5pt](1,1){.5}{0}{180}
\psarc[linecolor=blue,linewidth=1.5pt](1,3){.5}{180}{0}
\rput(4.57,2){\color{blue}$\dots$}
\end{pspicture}
\ +
\begin{pspicture}[.45](-.25,.75)(6,3)
\conn{(0,1)}{(6,3)}
\psline[linecolor=blue,linewidth=1.5pt](0.5,1)(0.5,3)
\psline[linecolor=blue,linewidth=1.5pt](3.5,1)(3.5,3)
\psline[linecolor=blue,linewidth=1.5pt](5.5,1)(5.5,3)
\psarc[linecolor=blue,linewidth=1.5pt](2,1){.5}{0}{180}
\psarc[linecolor=blue,linewidth=1.5pt](2,3){.5}{180}{0}
\rput(4.57,2){\color{blue}$\dots$}
\end{pspicture}
\ +\ \dots\ +
\begin{pspicture}[.45](-.25,.75)(6,3)
\conn{(0,1)}{(6,3)}
\psline[linecolor=blue,linewidth=1.5pt](0.5,1)(0.5,3)
\psline[linecolor=blue,linewidth=1.5pt](1.5,1)(1.5,3)
\psline[linecolor=blue,linewidth=1.5pt](3.5,1)(3.5,3)
\psarc[linecolor=blue,linewidth=1.5pt](5,1){.5}{0}{180}
\psarc[linecolor=blue,linewidth=1.5pt](5,3){.5}{180}{0}
\rput(2.57,2){\color{blue}$\dots$}
\end{pspicture}
\label{H}
\ee
which in terms of the generators of the {\em linear} TL algebra 
merely corresponds to
\be
  \Hb\ =\ -\sum_{j=1}^{N-1}e_j
\label{Hlin}
\ee
The set of normalized eigenvectors of $\Hb$ is the same
as that of $\Db(u)$ while the eigenvalues of $\Hb$ and $\Db(u)$ are different.
The {\em conformal} spectra are nevertheless the same as we will see below.

As preparation for the discussion of fusion in Section~7,
it is useful to adjust this definition of $\Hb$ to the various sectors.
We thus introduce the $s$-dependent Hamiltonian $\Hb_{(1,s)}$ by augmenting
the $N$-tangles in (\ref{H}) by the identity $(s-1)$-tangle, that is,
\psset{unit=.6cm}
\setlength{\unitlength}{.6cm}
\be
  \Hb_{(1,s)}\ =\ \begin{pspicture}[.45](-.25,.5)(10,3.6)
\conn{(0,1)}{(10,3)}
\rput(3.3,1.9){$\Hb_{(1,1)}$}
\psline[linecolor=black,linewidth=1pt,linestyle=dashed,dash=.25 
.25](6,.7)(6,3.3)
\psline[linecolor=blue,linewidth=1.5pt](6.5,1)(6.5,3)
\psline[linecolor=blue,linewidth=1.5pt](7.5,1)(7.5,3)
\psline[linecolor=blue,linewidth=1.5pt](9.5,1)(9.5,3)
\rput(8.57,2){\color{blue}$\dots$}
\end{pspicture}
\label{Hs}
\ee
where $\Hb_{(1,1)}=\Hb$. This augmented Hamiltonian acts naturally on
link states with $N+s-1$ nodes (and zero defects) of which the $s-1$ 
right-most nodes,
called {\em boundary nodes}, must be connected to nodes of the original
$N$-tangle Hamiltonian, called {\em bulk nodes}.
Likewise, a resulting state where
a pair of boundary nodes are connected, is set to zero. The vertical 
dashed line
in (\ref{Hs}) indicates the separation into bulk and boundary parts.
The spectrum of $\Hb_{(1,s)}$ is identical to the spectrum of $\Hb$ when the
latter is restricted to acting on the $(1,s)$ sector.

We will renormalize the Hamiltonian $\Hb_{(1,s)}$ by shifting the 
`ground state energy'
to facilitate the introduction of the finitized dilatation generator 
and to make the relation to the
finitized characters more transparent. This also ensures
that the renormalized Hamiltonian $\Hc_{(1,s)}$ is non-negative definite.
We define it by
\be
  \Hc_{(1,s)}\ =\ 
\Hb_{(1,s)}+2\big(\sum_{j=1}^{\lfloor\frac{N-1}{2}\rfloor}\sin 
t_j\big) \Ib
  \ =\ \Hb_{(1,s)}+   \begin{cases}\displaystyle
 
\frac{\sqrt{2}\sin\frac{(N-2)\pi}{4N}}{\sin\frac{\pi}{2N}}\Ib,&\mbox{$s$ 
odd}\\[18pt]
 
\displaystyle\frac{2\sin^2\frac{(N-1)\pi}{4N}}{\sin\frac{\pi}{2N}}\Ib,&\mbox{$s$ 
even}
  \end{cases}
\label{Hc1s}
\ee
where $t_j$ is defined in (\ref{tj}), and find that its eigenvalues 
are given by
\be
   2\sum_{j\in{\cal E}^{(s)}_N}\sin t_j+2\sum_{j\in{\cal M}^{(s)}_N}\sin t_j
\label{Hc1seigen}
\ee

Since the eigenvalues of $\Hc_{(1,s)}$ can be expressed as the linear
combinations (\ref{Hc1seigen}) of terms like $\sin t_j$, we may now define
the {\em finitized dilatation
generator} $L_0^{(1,s)}$ by replacing these summands $\sin t_j$ by $Nt_j/2\pi$
in the Jordan canonical form of $\Hc_{(1,s)}-\frac{1+(-1)^s}{16}\Ib$ 
when the diagonal
entries of the latter are expressed as in (\ref{Hc1seigen}). That is, 
we are defining
the finitized dilatation generator by the formal replacement
\be
  L_0^{(1,s)}\ =\
  {\rm Jordan}\Big(\Hc_{(1,s)}-\frac{1+(-1)^s}{16}\Ib\Big)\Bigg|_{\sin 
t_j\mapsto \frac{Nt_j}{2\pi}}
\label{L0subs}
\ee
The dependence on $N$ has been suppressed.
An eventual off-diagonal part of this Jordan canonical form is 
unaltered by this replacement.
We observe, though, that the Hamiltonian $\Hc_{(1,s)}$ is {\em diagonalizable}.
This is in accordance with our observation in Section~3.3 that the 
double-row transfer matrix
$\Db(u)$ itself is diagonalizable when restricted to the sector $(1,s)$.
In the discussion of fusion in Section~7, on the other hand, we will 
encounter Hamiltonians
which are {\em not} diagonalizable.

It also follows that the eigenvalues of $L_0^{(1,s)}$ are
\bea
   \sum_{j\in{\cal E}^{(s)}_N}j+\sum_{j\in{\cal M}^{(s)}_N}j, 
\hspace{1.4cm}&&\mbox{$s$ odd}\nn
   -\frac{1}{8}+\sum_{j\in{\cal E}^{(s)}_N}(j-\half)
      +\sum_{j\in{\cal M}^{(s)}_N}(j-\half), &&\mbox{$s$ even}
\label{L0}
\eea
in accordance with (\ref{logLafin}). This confirms the assertion 
above that the double-row transfer matrix
and the Hamiltonian have the same conformal spectra.
We can also re-express the finitized characters as
\be
  \chi_{1,s}^{(N)}(q)\ =\ \mathop{\rm Tr} q^{L_0^{(1,s)}-c/24}
\label{L0chi}
\ee
thus mimicking the definition of the ordinary Virasoro characters.

\section{Fusion}

In this section, we explain the diagrammatic implementation of fusion 
within our framework~\cite{PRZ}.
Although we focus here on critical dense polymers and the sectors 
with $r=1$ (labelled by $s=1,2,\ldots$),
the construction extends to the other logarithmic minimal models and 
sectors with $r>1$, albeit with more involved computations. For 
convenience in the description of our fusion prescription,
we work with the Hamiltonian limit of the double-row transfer matrix.
The conclusions about fusion are the same but more tedious to reach
when the analysis is based on the double-row transfer matrix $\Db(u)$ itself.

\subsection{Fusion prescription}

To study the fusion of the representations
$(1,s)$ and $(1,s')$, we consider a system of size $N$ augmented by
$s-1$ auxiliary nodes to the left and $s'-1$ auxiliary nodes to the right.
That is, the corresponding Hamiltonian is
\psset{unit=.6cm}
\setlength{\unitlength}{.6cm}
\be
  \Hb_{(1,s)|(1,s')}\ =\ \begin{pspicture}[.45](-.25,.5)(14,3.6)
\conn{(0,1)}{(14,3)}
\psline[linecolor=black,linewidth=1pt,linestyle=dashed,dash=.25 
.25](4,.7)(4,3.3)
\psline[linecolor=blue,linewidth=1.5pt](0.5,1)(0.5,3)
\psline[linecolor=blue,linewidth=1.5pt](1.5,1)(1.5,3)
\psline[linecolor=blue,linewidth=1.5pt](3.5,1)(3.5,3)
\rput(2.57,2){\color{blue}$\dots$}
\rput(7.3,1.9){$\Hb_{(1,1)}$}
\psline[linecolor=black,linewidth=1pt,linestyle=dashed,dash=.25 
.25](10,.7)(10,3.3)
\psline[linecolor=blue,linewidth=1.5pt](10.5,1)(10.5,3)
\psline[linecolor=blue,linewidth=1.5pt](11.5,1)(11.5,3)
\psline[linecolor=blue,linewidth=1.5pt](13.5,1)(13.5,3)
\rput(12.57,2){\color{blue}$\dots$}
\end{pspicture}
\label{Hss}
\ee
where $N\geq s+s'-2$ and $N-s-s'=0$ mod $2$. It acts naturally on 
link states with
$N+s+s'-2$ nodes and zero defects, though not all such link states are allowed.
Our fusion prescription {\em excludes} the link states where a pair of
nodes from the {\em same} boundary are linked. This means, in particular,
that a left-boundary node must be linked to either a right-boundary 
node or a bulk node
but not to another left-boundary node. Suppressing the dependence on $N$,
this set of link states is denoted ${\cal L}_{(1,s)|(1,s')}$.

Letting $h$ denote the number of half-arcs connecting nodes on the 
left boundary with nodes
on the right boundary, the space ${\cal L}_{(1,s)|(1,s')}$ is 
naturally decomposed as
the disjoint union of the spaces ${\cal L}_{(1,s)|(1,s')}^{(h)}$ 
containing link states
with fixed $h$, that is,
\be
  {\cal L}_{(1,s)|(1,s')}\ =\ {\cal 
L}_{(1,s)|(1,s')}^{(\min(s,s')-1)}\cup \dots\cup
  {\cal L}_{(1,s)|(1,s')}^{(1)} \cup {\cal L}_{(1,s)|(1,s')}^{(0)}
\label{LLL}
\ee
Viewed from the bulk, links to the left boundary are equivalent to links
to the right boundary as they are all just defects.
We may therefore re-interpret the states in ${\cal L}_{(1,s)|(1,s')}^{(h)}$ as
states in ${\cal L}_{(1,1)|(1,s+s'-1-2h)}^{(0)}={\cal L}_{(1,s+s'-1-2h)}$.
Explicitly, for the case $(1,4)|(1,3)$ with $N=7$ and $h=1$, we can 
map state-by-state as follows. First we move the first node from the 
left to the right boundary carrying its link (shown dashed and in 
blue in the illustration
(\ref{reint}) below) with it in a planar fashion. We repeat this with 
the second and third nodes on the left so that now all boundary nodes 
are on the right. Finally, we remove the $h$ spectator half-arcs (in 
this
case, the single one shown dashed and in blue) on the right boundary 
which are common to all allowed link states. This two-step
process is illustrated here
\psset{unit=.55cm}
\setlength{\unitlength}{.55cm}
\be
\begin{pspicture}[.25](0,-.95)(6,2.8)
\psarc[linecolor=blue,linewidth=1.5pt,linestyle=dashed,dash=.2 
.075](3,-0.75){2.75}{0}{180}
\psarc[linecolor=Maroon,linewidth=1.5pt](2,-0.75){1.25}{0}{180}
\psarc[linecolor=Maroon,linewidth=1.5pt](1.5,-0.75){.25}{0}{180}
\psarc[linecolor=Maroon,linewidth=1.5pt](2.5,-0.75){.25}{0}{180}
\psarc[linecolor=Maroon,linewidth=1.5pt](4.5,-0.75){.25}{0}{180}
\psarc[linecolor=Maroon,linewidth=1.5pt](4.5,-0.75){.75}{0}{180}
\psline[linecolor=black,linewidth=1pt,linestyle=dashed,dash=.25 
.25](1.5,-1)(1.5,2.25)
\psline[linecolor=black,linewidth=1pt,linestyle=dashed,dash=.25 
.25](5,-1)(5,2.25)
\end{pspicture}\quad \leftrightarrow\quad
\begin{pspicture}[.25](0,-.95)(6,2.8)
\psarc[linecolor=Maroon,linewidth=1.5pt](3,-0.75){2.75}{0}{180}
\psarc[linecolor=Maroon,linewidth=1.5pt](1,-0.75){.25}{0}{180}
\psarc[linecolor=Maroon,linewidth=1.5pt](3.5,-0.75){1.75}{0}{180}
\psarc[linecolor=blue,linewidth=1.5pt,linestyle=dashed,dash=.15 
.05](4.5,-0.75){.25}{0}{180}
\psarc[linecolor=Maroon,linewidth=1.5pt](3,-0.75){.25}{0}{180}
\psarc[linecolor=Maroon,linewidth=1.5pt](3,-0.75){.75}{0}{180}
\psline[linecolor=black,linewidth=1pt,linestyle=dashed,dash=.25 
.25](0,-1)(0,2.25)
\psline[linecolor=black,linewidth=1pt,linestyle=dashed,dash=.25 
.25](3.5,-1)(3.5,2.25)
\end{pspicture}
\quad \leftrightarrow\quad
\begin{pspicture}[.25](0,-.95)(5,2.8)
\psarc[linecolor=Maroon,linewidth=1.5pt](2.5,-0.75){2.25}{0}{180}
\psarc[linecolor=Maroon,linewidth=1.5pt](1,-0.75){.25}{0}{180}
\psarc[linecolor=Maroon,linewidth=1.5pt](3,-0.75){1.25}{0}{180}
\psarc[linecolor=Maroon,linewidth=1.5pt](3,-0.75){.25}{0}{180}
\psarc[linecolor=Maroon,linewidth=1.5pt](3,-0.75){.75}{0}{180}
\psline[linecolor=black,linewidth=1pt,linestyle=dashed,dash=.25 
.25](0,-1)(0,2.25)
\psline[linecolor=black,linewidth=1pt,linestyle=dashed,dash=.25 
.25](3.5,-1)(3.5,2.25)
\end{pspicture}
\label{reint}
\ee

This re-interpretation naively implies the $sl(2)$ fusion rule
\be
(1,s_1)\otimes_f (1,s_2)\ =\ \bigoplus_s\ (1,s)
\label{sl2}
\ee
where the sum is over $s=|s_1-s_2|+1,|s_1-s_2|+3,\ldots,s_1+s_2-1$.
Care must be exercised, though, since defects can be annihilated in pairs.
Indeed, this latter property implies that the action of the 
Hamiltonian on (\ref{LLL})
is upper {\em block-triangular} rather than block-diagonal.
There is thus the possibility of forming {\em indecomposable representations}
in which case the right side of (\ref{sl2}) is {\em not} a direct sum.
In the following, we will characterize these representations and conjecture
the fusion rules applying to the complete set of representations generated
by fusion of the $(1,s)$ building blocks.

Due to the decomposition (\ref{LLL}), the renormalization of the Hamiltonian
$\Hb_{(1,s)|(1,s')}$ is defined as in (\ref{Hc1s}) by
\be
  \Hc_{(1,s)|(1,s')}\ =\ \Hb_{(1,s)|(1,s')}
   +  \begin{cases}\displaystyle
   \frac{\sqrt{2}\sin\frac{(N-2)\pi}{4N}}{\sin\frac{\pi}{2N}}\Ib,\ \ \ 
&\mbox{$s+s'$ even}\\[18pt]
 
\displaystyle\frac{2\sin^2\frac{(N-1)\pi}{4N}}{\sin\frac{\pi}{2N}}\Ib,&\mbox{$s+s'$ 
odd}
  \end{cases}
\label{Hcss}
\ee
Due to the decomposition (\ref{LLL}) and the subsequent 
re-interpretation, the eigenvalues
of $\Hc_{(1,s)|(1,s')}$ are
\be
   2\!\!\!\!\!\sum_{j\in{\cal E}^{(s+s'-1-2h)}_N}\!\!\!\sin t_j
    +2\!\!\!\!\!\!\sum_{j\in{\cal M}^{(s+s'-1-2h)}_N}\!\!\!\sin t_j,\ 
\ \ \ \ \ \
     h=0,\ldots,\min(s,s')-1
\label{Hcsseigen}
\ee
The associated finitized dilatation generator thus reads
\be
  L_0^{(1,s)|(1,s')}\ =\
   {\rm Jordan}\Big(\Hc_{(1,s)|(1,s')}-\frac{1-(-1)^{s+s'}}{16}\Ib
     \Big)\Bigg|_{\sin t_j\mapsto \frac{Nt_j}{2\pi}}
\label{L0ss}
\ee
and has eigenvalues
\bea
   \sum_{j\in{\cal E}^{(s+s'-1-2h)}_N}\!\!\!j+\!\!\!\!\sum_{j\in{\cal 
M}^{(s+s'-1-2h)}_N}\!\!\!\!j,
   \qquad\hspace{1.4cm} h=0,\ldots,\min(s,s')-1,&& \mbox{$s+s'$ even}\nn
   -\frac{1}{8}+\!\!\!\sum_{j\in{\cal E}^{(s+s'-1-2h)}_N}\!\!\!(j-\half)
      +\!\!\!\!\sum_{j\in{\cal M}^{(s+s'-1-2h)}_N}\!\!\!\!(j-\half),
    \qquad h=0,\ldots,\min(s,s')-1, && \mbox{$s+s'$ odd}
  \nn
\eea

\subsection{Indecomposable representations}

As a first example, we consider the fusion
\be
  (1,2)\otimes_f(1,2)\ =\ (1,1)\oplus_i(1,3)
\label{1212}
\ee
in some detail. The subscript $i$ refers to the indecomposable
structure of the right side. Let the system size be $N=4$ in which case
\psset{unit=.48cm}
\setlength{\unitlength}{.48cm}
\bea
  {\cal L}_{(1,2)|(1,2)}^{(1)}&=&\Big\{
  \begin{pspicture}(3,1.5)
\psarc[linecolor=Maroon,linewidth=1.5pt](1,-0.5){.25}{0}{180}
\psarc[linecolor=Maroon,linewidth=1.5pt](1.5,-0.5){1.25}{0}{180}
\psarc[linecolor=Maroon,linewidth=1.5pt](2,-0.5){.25}{0}{180}
\psline[linecolor=black,linewidth=1pt,linestyle=dashed,dash=.25 
.25](0.5,-.75)(0.5,1.05)
\psline[linecolor=black,linewidth=1pt,linestyle=dashed,dash=.25 
.25](2.5,-.75)(2.5,1.05)
\rput(3.02,-0.5){,}
\end{pspicture}\
\begin{pspicture}(3,1.5)
\psarc[linecolor=Maroon,linewidth=1.5pt](1.5,-0.5){.25}{0}{180}
\psarc[linecolor=Maroon,linewidth=1.5pt](1.5,-0.5){.75}{0}{180}
\psarc[linecolor=Maroon,linewidth=1.5pt](1.5,-0.5){1.25}{0}{180}
\psline[linecolor=black,linewidth=1pt,linestyle=dashed,dash=.25 
.25](0.5,-.75)(0.5,1.05)
\psline[linecolor=black,linewidth=1pt,linestyle=dashed,dash=.25 
.25](2.5,-.75)(2.5,1.05)
\end{pspicture}
      \Big\}\nn
  {\cal L}_{(1,2)|(1,2)}^{(0)}&=&\big\{
  \begin{pspicture}(3,1.5)
\psarc[linecolor=Maroon,linewidth=1.5pt](0.5,-0.25){.25}{0}{180}
\psarc[linecolor=Maroon,linewidth=1.5pt](1.5,-0.25){.25}{0}{180}
\psarc[linecolor=Maroon,linewidth=1.5pt](2.5,-0.25){.25}{0}{180}
\psline[linecolor=black,linewidth=1pt,linestyle=dashed,dash=.25 
.25](0.5,-.5)(0.5,.75)
\psline[linecolor=black,linewidth=1pt,linestyle=dashed,dash=.25 
.25](2.5,-.5)(2.5,.75)
\rput(3.02,-0.25){,}
\end{pspicture}\
  \begin{pspicture}(3,1.5)
\psarc[linecolor=Maroon,linewidth=1.5pt](1,-0.25){.25}{0}{180}
\psarc[linecolor=Maroon,linewidth=1.5pt](1,-0.25){.75}{0}{180}
\psarc[linecolor=Maroon,linewidth=1.5pt](2.5,-0.25){.25}{0}{180}
\psline[linecolor=black,linewidth=1pt,linestyle=dashed,dash=.25 
.25](0.5,-.5)(0.5,.75)
\psline[linecolor=black,linewidth=1pt,linestyle=dashed,dash=.25 
.25](2.5,-.5)(2.5,.75)
\rput(3.02,-0.25){,}
\end{pspicture}\
  \begin{pspicture}(3,1.5)
\psarc[linecolor=Maroon,linewidth=1.5pt](0.5,-0.25){.25}{0}{180}
\psarc[linecolor=Maroon,linewidth=1.5pt](2,-0.25){.25}{0}{180}
\psarc[linecolor=Maroon,linewidth=1.5pt](2,-0.25){.75}{0}{180}
\psline[linecolor=black,linewidth=1pt,linestyle=dashed,dash=.25 
.25](0.5,-.5)(0.5,.75)
\psline[linecolor=black,linewidth=1pt,linestyle=dashed,dash=.25 
.25](2.5,-.5)(2.5,.75)
\end{pspicture}
  \big\}
\label{L1212}
\eea
Acting on these link states listed in the order indicated here, the 
matrix representation
of the Hamiltonian $\Hb_{(1,2)|(1,2)}$ is
\be
   \Hb_{(1,2)|(1,2)}\ =\ -
   \begin{pmatrix} 0&2&0&1&1\\ 1&0&0&0&0\\ 0&0&0&1&1\\
    0&0&1&0&0\\ 0&0&1&0&0 \end{pmatrix}
\label{H1212}
\ee
The Jordan canonical form of the Hamiltonian $\Hc_{(1,2)|(1,2)}$
thus reads
\be
  {\rm Jordan}\big(\Hc_{(1,2)|(1,2)}\big)\ =\
   {\rm diag}\Big[ \begin{pmatrix} 0&1\\ 0&0 \end{pmatrix},2\sin\frac{\pi}{4},
    \begin{pmatrix} 4\sin\frac{\pi}{4}&1\\ 0&4\sin\frac{\pi}{4} 
\end{pmatrix} \Big]
\label{Hc1212}
\ee
in which case the finitized dilatation generator, after the 
replacement $\sin\frac{\pi}{4}\mapsto 
\frac{4\frac{\pi}{4}}{2\pi}=\half$, becomes
\be
  L_0^{(1,2)|(1,2)}\ =\
   {\rm diag}\Big[\begin{pmatrix} 0&1\\ 0&0 \end{pmatrix},1,
    \begin{pmatrix} 2&1\\ 0&2 \end{pmatrix} \Big]
\label{L01212}
\ee
Also for $N=4$, the finitized dilatation generators associated to
$\Hb_{(1,1)}$ and $\Hb_{(1,3)}$ are likewise found to be
\be
  L_0^{(1,1)}\ =\ {\rm diag}[0,2],\ \ \ \ \ \ \ L_0^{(1,3)}\ =\ {\rm 
diag}[0,1,2]
\ee
This means that
the finitized partition function associated to (\ref{H1212}) decomposes as
\bea
  Z^{(4)}_{(1,2)|(1,2)}(q)\ =\ \chi^{(4)}_{(1,1)}(q)+\chi^{(4)}_{(1,3)}(q)
   &=&q^{1/12}[(1+q^2)+(1+q+q^2)]\nn
   &=&q^{1/12}(2+q+2q^2)
\label{Z1212}
\eea
in accordance with the fusion rule (\ref{1212}).

The decomposition (\ref{Z1212}) does not contain information on the 
rank-two Jordan cells
appearing in (\ref{L01212}) as the characters only reflect the diagonal part.
The non-trivial Jordan-cell structure is a finite-size manifestation of the
right side of (\ref{1212}) being {\em indecomposable}.
It is observed that a Jordan cell is formed {\em whenever possible} and
we have verified for $N=2,4,6,8$ that this property persists.
It is also noted that the Hamiltonian is {\em upper}
block-triangular meaning that states contributing to $\chit_3(q)$ can 
be mapped to
states contributing to $\chit_1(q)$ but not vice versa.
These observations are compatible with the properties of the 
indecomposable representation
${\cal R}_{1,1}$, appearing in~\cite{GK}, with character 
$\chit_1(q)+\chit_3(q)$.
Assuming equivalence of the indecomposable representations,
we will adopt their notation here and henceforth denote
the right side of (\ref{1212}) by ${\cal R}_{1,1}$.

As a second example, we consider the fusion
\be
  (1,2)\otimes_f(1,4)\ =\ (1,3)\oplus_i(1,5)
\label{1214}
\ee
Let the system size be $N=6$ in which case
\psset{unit=.48cm}
\setlength{\unitlength}{.48cm}
\bea
  {\cal L}_{(1,2)|(1,4)}^{(1)}&=&\Big\{
  \begin{pspicture}(5,2.5)
\psarc[linecolor=Maroon,linewidth=1.5pt](1,-0.75){.25}{0}{180}
\psarc[linecolor=Maroon,linewidth=1.5pt](2,-0.75){.25}{0}{180}
\psarc[linecolor=Maroon,linewidth=1.5pt](2.5,-0.75){2.25}{0}{180}
\psarc[linecolor=Maroon,linewidth=1.5pt](3.5,-0.75){.25}{0}{180}
\psarc[linecolor=Maroon,linewidth=1.5pt](3.5,-0.75){.75}{0}{180}
\psline[linecolor=black,linewidth=1pt,linestyle=dashed,dash=.25 
.25](0.5,-1)(0.5,1.7)
\psline[linecolor=black,linewidth=1pt,linestyle=dashed,dash=.25 
.25](3.5,-1)(3.5,1.7)
\rput(5.02,-0.75){,}
\end{pspicture}\
  \begin{pspicture}(5,2.5)
\psarc[linecolor=Maroon,linewidth=1.5pt](1.5,-0.75){.25}{0}{180}
\psarc[linecolor=Maroon,linewidth=1.5pt](1.5,-0.75){.75}{0}{180}
\psarc[linecolor=Maroon,linewidth=1.5pt](2.5,-0.75){2.25}{0}{180}
\psarc[linecolor=Maroon,linewidth=1.5pt](3.5,-0.75){.25}{0}{180}
\psarc[linecolor=Maroon,linewidth=1.5pt](3.5,-0.75){.75}{0}{180}
\psline[linecolor=black,linewidth=1pt,linestyle=dashed,dash=.25 
.25](0.5,-1)(0.5,1.7)
\psline[linecolor=black,linewidth=1pt,linestyle=dashed,dash=.25 
.25](3.5,-1)(3.5,1.7)
\rput(5.02,-0.75){,}
\end{pspicture}\
  \begin{pspicture}(5,2.5)
\psarc[linecolor=Maroon,linewidth=1.5pt](1,-0.75){.25}{0}{180}
\psarc[linecolor=Maroon,linewidth=1.5pt](2.5,-0.75){.25}{0}{180}
\psarc[linecolor=Maroon,linewidth=1.5pt](2.5,-0.75){2.25}{0}{180}
\psarc[linecolor=Maroon,linewidth=1.5pt](3.5,-0.75){.25}{0}{180}
\psarc[linecolor=Maroon,linewidth=1.5pt](3,-0.75){1.25}{0}{180}
\psline[linecolor=black,linewidth=1pt,linestyle=dashed,dash=.25 
.25](0.5,-1)(0.5,1.7)
\psline[linecolor=black,linewidth=1pt,linestyle=dashed,dash=.25 
.25](3.5,-1)(3.5,1.7)
\rput(5.02,-0.75){,}
\end{pspicture}\
  \begin{pspicture}(5,2.5)
\psarc[linecolor=Maroon,linewidth=1.5pt](1,-0.75){.25}{0}{180}
\psarc[linecolor=Maroon,linewidth=1.5pt](3,-0.75){.25}{0}{180}
\psarc[linecolor=Maroon,linewidth=1.5pt](2.5,-0.75){2.25}{0}{180}
\psarc[linecolor=Maroon,linewidth=1.5pt](3,-0.75){1.25}{0}{180}
\psarc[linecolor=Maroon,linewidth=1.5pt](3,-0.75){.75}{0}{180}
\psline[linecolor=black,linewidth=1pt,linestyle=dashed,dash=.25 
.25](0.5,-1)(0.5,1.7)
\psline[linecolor=black,linewidth=1pt,linestyle=dashed,dash=.25 
.25](3.5,-1)(3.5,1.7)
\rput(5.02,-0.75){,}
\end{pspicture}\nn
&&\hspace{-.2cm}
  \begin{pspicture}(5,2.5)
\psarc[linecolor=Maroon,linewidth=1.5pt](2.5,-0.75){.25}{0}{180}
\psarc[linecolor=Maroon,linewidth=1.5pt](2.5,-0.75){1.75}{0}{180}
\psarc[linecolor=Maroon,linewidth=1.5pt](2.5,-0.75){2.25}{0}{180}
\psarc[linecolor=Maroon,linewidth=1.5pt](1.5,-0.75){.25}{0}{180}
\psarc[linecolor=Maroon,linewidth=1.5pt](3.5,-0.75){.25}{0}{180}
\psline[linecolor=black,linewidth=1pt,linestyle=dashed,dash=.25 
.25](0.5,-1)(0.5,1.7)
\psline[linecolor=black,linewidth=1pt,linestyle=dashed,dash=.25 
.25](3.5,-1)(3.5,1.7)
\rput(5.02,-0.75){,}
\end{pspicture}\
  \begin{pspicture}(5,2.5)
\psarc[linecolor=Maroon,linewidth=1.5pt](1.5,-0.75){.25}{0}{180}
\psarc[linecolor=Maroon,linewidth=1.5pt](2.5,-0.75){1.75}{0}{180}
\psarc[linecolor=Maroon,linewidth=1.5pt](2.5,-0.75){2.25}{0}{180}
\psarc[linecolor=Maroon,linewidth=1.5pt](3,-0.75){.25}{0}{180}
\psarc[linecolor=Maroon,linewidth=1.5pt](3,-0.75){.75}{0}{180}
\psline[linecolor=black,linewidth=1pt,linestyle=dashed,dash=.25 
.25](0.5,-1)(0.5,1.7)
\psline[linecolor=black,linewidth=1pt,linestyle=dashed,dash=.25 
.25](3.5,-1)(3.5,1.7)
\rput(5.02,-0.75){,}
\end{pspicture}\
\begin{pspicture}(5,2.5)
\psarc[linecolor=Maroon,linewidth=1.5pt](2,-0.75){.25}{0}{180}
\psarc[linecolor=Maroon,linewidth=1.5pt](2.5,-0.75){1.75}{0}{180}
\psarc[linecolor=Maroon,linewidth=1.5pt](2.5,-0.75){2.25}{0}{180}
\psarc[linecolor=Maroon,linewidth=1.5pt](2,-0.75){.75}{0}{180}
\psarc[linecolor=Maroon,linewidth=1.5pt](3.5,-0.75){.25}{0}{180}
\psline[linecolor=black,linewidth=1pt,linestyle=dashed,dash=.25 
.25](0.5,-1)(0.5,1.7)
\psline[linecolor=black,linewidth=1pt,linestyle=dashed,dash=.25 
.25](3.5,-1)(3.5,1.7)
\rput(5.02,-0.75){,}
\end{pspicture}\
\begin{pspicture}(5,2.5)
\psarc[linecolor=Maroon,linewidth=1.5pt](2.5,-0.75){1.25}{0}{180}
\psarc[linecolor=Maroon,linewidth=1.5pt](2.5,-0.75){1.75}{0}{180}
\psarc[linecolor=Maroon,linewidth=1.5pt](2.5,-0.75){2.25}{0}{180}
\psarc[linecolor=Maroon,linewidth=1.5pt](2,-0.75){.25}{0}{180}
\psarc[linecolor=Maroon,linewidth=1.5pt](3,-0.75){.25}{0}{180}
\psline[linecolor=black,linewidth=1pt,linestyle=dashed,dash=.25 
.25](0.5,-1)(0.5,1.7)
\psline[linecolor=black,linewidth=1pt,linestyle=dashed,dash=.25 
.25](3.5,-1)(3.5,1.7)
\rput(5.02,-0.75){,}
\end{pspicture}\
\begin{pspicture}(5,2.5)
\psarc[linecolor=Maroon,linewidth=1.5pt](2.5,-0.75){.25}{0}{180}
\psarc[linecolor=Maroon,linewidth=1.5pt](2.5,-0.75){1.75}{0}{180}
\psarc[linecolor=Maroon,linewidth=1.5pt](2.5,-0.75){2.25}{0}{180}
\psarc[linecolor=Maroon,linewidth=1.5pt](2.5,-0.75){.75}{0}{180}
\psarc[linecolor=Maroon,linewidth=1.5pt](2.5,-0.75){1.25}{0}{180}
\psline[linecolor=black,linewidth=1pt,linestyle=dashed,dash=.25 
.25](0.5,-1)(0.5,1.7)
\psline[linecolor=black,linewidth=1pt,linestyle=dashed,dash=.25 
.25](3.5,-1)(3.5,1.7)
\end{pspicture}
      \Big\}\nn
  {\cal L}_{(1,2)|(1,4)}^{(0)}&=&\Big\{
  \begin{pspicture}(5,2.5)
\psarc[linecolor=Maroon,linewidth=1.5pt](.5,-0.5){.25}{0}{180}
\psarc[linecolor=Maroon,linewidth=1.5pt](1.5,-0.5){.25}{0}{180}
\psarc[linecolor=Maroon,linewidth=1.5pt](3.5,-0.5){.25}{0}{180}
\psarc[linecolor=Maroon,linewidth=1.5pt](3.5,-0.5){.75}{0}{180}
\psarc[linecolor=Maroon,linewidth=1.5pt](3.5,-0.5){1.25}{0}{180}
\psline[linecolor=black,linewidth=1pt,linestyle=dashed,dash=.25 
.25](0.5,-1)(0.5,1.7)
\psline[linecolor=black,linewidth=1pt,linestyle=dashed,dash=.25 
.25](3.5,-1)(3.5,1.7)
\rput(5.02,-0.5){,}
\end{pspicture}\
  \begin{pspicture}(5,2.5)
\psarc[linecolor=Maroon,linewidth=1.5pt](1,-0.5){.25}{0}{180}
\psarc[linecolor=Maroon,linewidth=1.5pt](1,-0.5){.75}{0}{180}
\psarc[linecolor=Maroon,linewidth=1.5pt](3.5,-0.5){1.25}{0}{180}
\psarc[linecolor=Maroon,linewidth=1.5pt](3.5,-0.5){.25}{0}{180}
\psarc[linecolor=Maroon,linewidth=1.5pt](3.5,-0.5){.75}{0}{180}
\psline[linecolor=black,linewidth=1pt,linestyle=dashed,dash=.25 
.25](0.5,-1)(0.5,1.7)
\psline[linecolor=black,linewidth=1pt,linestyle=dashed,dash=.25 
.25](3.5,-1)(3.5,1.7)
\rput(5.02,-0.5){,}
\end{pspicture}\
  \begin{pspicture}(5,2.5)
\psarc[linecolor=Maroon,linewidth=1.5pt](.5,-0.5){.25}{0}{180}
\psarc[linecolor=Maroon,linewidth=1.5pt](2,-0.5){.25}{0}{180}
\psarc[linecolor=Maroon,linewidth=1.5pt](3.5,-0.5){.75}{0}{180}
\psarc[linecolor=Maroon,linewidth=1.5pt](3.5,-0.5){.25}{0}{180}
\psarc[linecolor=Maroon,linewidth=1.5pt](3,-0.5){1.75}{0}{180}
\psline[linecolor=black,linewidth=1pt,linestyle=dashed,dash=.25 
.25](0.5,-1)(0.5,1.7)
\psline[linecolor=black,linewidth=1pt,linestyle=dashed,dash=.25 
.25](3.5,-1)(3.5,1.7)
\rput(5.02,-0.5){,}
\end{pspicture}\
  \begin{pspicture}(5,2.5)
\psarc[linecolor=Maroon,linewidth=1.5pt](.5,-0.5){.25}{0}{180}
\psarc[linecolor=Maroon,linewidth=1.5pt](2.5,-0.5){.25}{0}{180}
\psarc[linecolor=Maroon,linewidth=1.5pt](3.5,-0.5){.25}{0}{180}
\psarc[linecolor=Maroon,linewidth=1.5pt](3,-0.5){1.25}{0}{180}
\psarc[linecolor=Maroon,linewidth=1.5pt](3,-0.5){1.75}{0}{180}
\psline[linecolor=black,linewidth=1pt,linestyle=dashed,dash=.25 
.25](0.5,-1)(0.5,1.7)
\psline[linecolor=black,linewidth=1pt,linestyle=dashed,dash=.25 
.25](3.5,-1)(3.5,1.7)
\rput(5.02,-0.5){,}
\end{pspicture}\
  \begin{pspicture}(5,2.5)
\psarc[linecolor=Maroon,linewidth=1.5pt](.5,-0.5){.25}{0}{180}
\psarc[linecolor=Maroon,linewidth=1.5pt](3,-0.5){.25}{0}{180}
\psarc[linecolor=Maroon,linewidth=1.5pt](3,-0.5){.75}{0}{180}
\psarc[linecolor=Maroon,linewidth=1.5pt](3,-0.5){1.25}{0}{180}
\psarc[linecolor=Maroon,linewidth=1.5pt](3,-0.5){1.75}{0}{180}
\psline[linecolor=black,linewidth=1pt,linestyle=dashed,dash=.25 
.25](0.5,-1)(0.5,1.7)
\psline[linecolor=black,linewidth=1pt,linestyle=dashed,dash=.25 
.25](3.5,-1)(3.5,1.7)
\end{pspicture}
  \Big\}\nn
\label{L1214}
\eea
Acting on these link states listed in the order indicated here, the 
matrix representation
of the Hamiltonian $\Hb_{(1,2)|(1,4)}$ is
\be
\Hb_{(1,2)|(1,4)}\ =\ - \mbox{\scriptsize $\left(\begin{array}{cccccccccccccc}
   0&2&1&0&0&0&1&0&0&0&1&1&0&0  \\
   1&0&0&0&0&0&0&0&0&0&0&0&0&0  \\
   1&0&0&1&1&0&0&0&0&0&0&0&1&0  \\
   0&0&1&0&0&1&0&0&0&0&0&0&0&1  \\
   0&1&1&0&0&1&2&0&1&0&0&0&0&0  \\
   0&0&0&1&1&0&0&1&0&0&0&0&0&0  \\
   0&0&0&0&1&0&0&0&0&0&0&0&0&0  \\
   0&0&0&0&0&1&1&0&2&0&0&0&0&0  \\
   0&0&0&0&0&0&0&1&0&0&0&0&0&0  \\
   0&0&0&0&0&0&0&0&0& 0&1&1&0&0  \\
   0&0&0&0&0&0&0&0&0& 1&0&0&0&0  \\
   0&0&0&0&0&0&0&0&0& 1&0&0&1&0  \\
   0&0&0&0&0&0&0&0&0& 0&0&1&0&1  \\
   0&0&0&0&0&0&0&0&0& 0&0&0&1&0  \end{array}\right)$}
\label{H1214}
\ee
The Jordan canonical form of the associated renormalized Hamiltonian 
$\Hc_{(1,2)|(1,4)}$
thus reads
\bea
  \hspace{-.4cm}{\rm Jordan}\big(\Hc_{(1,2)|(1,4)}\big)&=&{\rm diag}\Big[
   0,\begin{pmatrix} 1&1\\ 0&1 \end{pmatrix},
   \begin{pmatrix} \sqrt{3}&1\\ 0&\sqrt{3} \end{pmatrix},2,
   1+\sqrt{3},1+\sqrt{3},2\sqrt{3},\nn
   &&
   \begin{pmatrix} 2+\sqrt{3}&1\\ 0&2+\sqrt{3} \end{pmatrix},
   \begin{pmatrix} 1+2\sqrt{3}&1\\ 0&1+2\sqrt{3} \end{pmatrix},
   2+2\sqrt{3}
  \Big]
\label{Hc1214}
\eea
where we have used that $2\sin\frac{\pi}{6}=1$ and 
$2\sin\frac{\pi}{3}=\sqrt{3}$.
The associated finitized dilatation generator is now found to be
\be
  L_0^{(1,2)|(1,4)}\ =\ {\rm diag}\Big[
   0,\begin{pmatrix} 1&1\\ 0&1 \end{pmatrix},
   \begin{pmatrix} 2&1\\ 0&2 \end{pmatrix},2,
   3,3,4,
   \begin{pmatrix} 4&1\\ 0&4 \end{pmatrix},
   \begin{pmatrix} 5&1\\ 0&5 \end{pmatrix},
   6
  \Big]
\label{L01214}
\ee
It is observed that the repeated eigenvalue 3 does {\em not} form a 
Jordan cell.
This is unlike the fusion (\ref{1212}) above where a Jordan cell is formed
whenever possible.
Also for $N=6$, the finitized dilatation generators associated to
$\Hb_{(1,3)}$ and $\Hb_{(1,5)}$ are likewise found to be
\be
  L_0^{(1,3)}\ =\ {\rm diag}[0,1,2,2,3,4,4,5,6],\ \ \ \ \ \ \ 
L_0^{(1,5)}\ =\ {\rm diag}[1,2,3,4,5]
\ee
This means that
the finitized partition function associated to (\ref{H1214}) decomposes as
\bea
  Z^{(6)}_{(1,2)|(1,4)}(q)&=&\chi^{(6)}_{(1,3)}(q)+\chi^{(6)}_{(1,5)}(q)\nn
   &=&q^{1/12}[(1+q+2q^2+q^3+2q^4+q^5+q^6)
    +q(1+q+q^2+q^3+q^4)]\nn
   &=&q^{1/12}(1+2q+3q^2+2q^3+3q^4+2q^5+q^6)
\label{Z1214}
\eea
and we note that the `missing' Jordan cell corresponds to $q^3$.
Continuing our comparison with the results in~\cite{GK}, we denote the
associated indecomposable representation $(1,3)\oplus_i(1,5)$
by ${\cal R}_{2,1}$.

Based on the examples above and many other explicit evaluations of
fusions of the form $(1,s)\otimes_f(1,s')$, we have observed that
indecomposable representations are only formed by the combinations
\be
{\cal R}_{j,1}\ =\  (1,2j-1)\oplus_i(1,2j+1),\qquad j=1,2,\ldots
\label{indecj}
\ee
and that they arise as the result of the fusions
\be
  (1,2)\otimes_f(1,2j)\ =\ (1,2j-1)\oplus_i(1,2j+1)
\label{1212j}
\ee
The characters and indecomposable structure of these indecomposable
representations correspond to the definition of ${\cal R}_{j,1}$ in~\cite{GK}.
In particular, the off-diagonal part of the Jordan decomposition of the
Hamiltonian or dilatation generator maps $(1,2j+1)$ to $(1,2j-1)$ but not the
other way. It is stressed that the Jordan cells rendering the 
representations $\Rc_{j,1}$
indecomposable all have rank two.
As we will argue in the following, no new indecomposable representations
arise when considering fusions of indecomposable representations.
The set $\{{\cal R}_{j,1}\}$ is also the complete set of 
indecomposable representations
appearing in~\cite{GK}.

Here we can refine the characterization of the indecomposable representation
${\cal R}_{j,1}$. Indeed, with reference to the fermionic expressions 
for the finitized
characters $\chit_{(1,2j-1)}^{(N)}(q)$ and 
$\chit_{(1,2j+1)}^{(N)}(q)$ in (\ref{ferm}),
we have found that in the cases examined,
\bea
    \chi^{(N)}_{{\cal 
R}_{j,1}}(q)&=&\chit_{(1,2j-1)}^{(N)}(q)+\chit_{(1,2j+1)}^{(N)}(q)\nn
&=&q^{\frac{1}{12}}\Big(\sum_{m=0}^{\frac{N-2j+2}{2}}\sbinlr{\frac{N-2}{2}}{m,m+j-2}_{\!q}
+\sum_{m=0}^{\frac{N-2j}{2}}\sbinlr{\frac{N-2}{2}}{m,m+j-1}_{\!q}\Big)\nonumber\\[.2cm]
&&\hspace{6cm}\nearrow\nn
   &&\hspace{1.5cm}
+q^{\frac{1}{12}}\Big(\sum_{m=0}^{\frac{N-2j}{2}}\sbinlr{\frac{N-2}{2}}{m,m+j-1}_{\!q}
   +\sum_{m=0}^{\frac{N-2j-2}{2}}\sbinlr{\frac{N-2}{2}}{m,m+j}_{\!q}\Big)
\label{Rj}
\eea
Here the arrow indicates the off-diagonal action of the dilatation generator
which maps every state in the given sum to its identical mirror state 
in the target sum.
No other off-diagonal action is present. Together with
the statement that the map is from $(1,2j+1)$ to $(1,2j-1)$ only, 
this gives a full description
of the indecomposable structure of (the finitized version of) ${\cal R}_{j,1}$.
It is noted that the number of Jordan cells appearing in the finitized version
of ${\cal R}_{j,1}$ is given by
\be
  \lim_{q\rightarrow1}\sum_{m=0}^{\frac{N-2j}{2}}\sbinlr{\frac{N-2}{2}}{m,m+j-1}_{\!q}
  \ =\ \bin{N-1}{\frac{N-2j}{2}}-\bin{N-1}{\frac{N+2j}{2}}  \ =\ 
\lim_{q\rightarrow1}\chi_{1,2j}^{(N-1)}(q)
\ee
We find it useful to view the indecomposable structure of ${\cal R}_{j,1}$
as the following formal four-dimensional matrix containing a {\em single}
rank-two Jordan cell
\bea
   &&\hspace{-1cm}{\rm diag}\Big[
   \sum_{m=0}^{\frac{N-2j+2}{2}}\sbinlr{\frac{N-2}{2}}{m,m+j-2}_{\!q},\nn
   &&\hspace{1.8cm}\begin{pmatrix}  \displaystyle
      \sum_{m=0}^{\frac{N-2j}{2}}\sbinlr{\frac{N-2}{2}}{m,m+j-1}_{\!q}&1\\
    0&\displaystyle 
\sum_{m=0}^{\frac{N-2j}{2}}\sbinlr{\frac{N-2}{2}}{m,m+j-1}_{\!q}
    \end{pmatrix},\nn
   &&\hspace{9cm} 
\sum_{m=0}^{\frac{N-2j-2}{2}}\sbinlr{\frac{N-2}{2}}{m,m+j}_{\!q}
   \Big]
\label{LI}
\eea

It follows and is emphasized that if a Jordan cell corresponding to a 
particular energy (that is, power of $q$ in the associated finitized 
character (\ref{Rj})) is present for a particular system size,
a Jordan cell corresponding to the {\em same} energy is present for 
{\em all\/} larger system sizes. This supports our claim that this 
particular Jordan-cell structure persists
in the limit $N\rightarrow\infty$.

Let us reexamine the two fusions $(1,2)\otimes_f(1,2)$ and 
$(1,2)\otimes_f(1,4)$ and
put into the perspective of (\ref{Rj}) why a Jordan cell was observed 
whenever possible
in the first of these fusions but not in the second.
For finite system size $N$, the right side of the fusion (\ref{1212}) 
is described by
(\ref{Rj}) with $j=1$. Since $\sbinlr{M}{m,n}$ vanishes if $m>n$, cf. 
(\ref{sbin}),
the sum 
$\sum_{m=0}^{\frac{N-2j+2}{2}}\sbinlr{\frac{N-2}{2}}{m,m+j-2}_{\!q}$ 
vanishes
and a Jordan cell is formed whenever possible.
Also for finite $N$, the right side of the fusion (\ref{1214}) is described by
(\ref{Rj}) with $j=2$. In this case, a Jordan cell is only formed for 
the matching sums
while the remaining two sums in general will involve states with 
equal energies.
This was illustrated in (\ref{Hc1214}) for $N=6$ in which case the 
two sums in question read
\be
\sum_{m=0}^{\frac{N-2j+2}{2}}\sbinlr{\frac{N-2}{2}}{m,m+j-2}_{\!q}\Bigg|_{N=6,j=2}\hspace{-.3in}=\; 
   1+q^2+q^3+q^4+q^6,\quad
\sum_{m=0}^{\frac{N-2j-2}{2}}\sbinlr{\frac{N-2}{2}}{m,m+j}_{\!q}\Bigg|_{N=6,j=2}\hspace{-.3in}=\;q^3
\label{j2}
\ee
This explains why the Jordan cell corresponding to $q^3$ was `missing'.

We wish to point out that one can be misled when considering very 
small system sizes $N$.
As this example illustrates
\be
  \chi_{\Rc_{j,1}}^{(2j-2)}(q)\ =\ q^{\half+\D_{1,2j-1}}\ =\ 
\chi_{(1,2j-1)}^{(2j-2)}(q)
\ee
an ambiguity due to finite-size effects can arise since the 
indecomposable structure
of $\Rc_{j,1}$ is only visible if $N$ is big enough to accommodate it.

\subsection{Fusion rules}

Our next task is to examine fusion of indecomposable representations.
We thus need a characterization of the boundary conditions corresponding
to ${\cal R}_{j,1}$. Since ${\cal R}_{j,1}=(1,2j-1)\oplus_i(1,2j+1)$, 
the boundary should
accommodate $2j-2$ as well as $2j$ boundary nodes. To preserve the number of
nodes in the associated link states, the former case is represented 
by $2j$ boundary nodes
of which two must be linked together by a spectator half-arc. The right
boundary associated to ${\cal R}_{j,1}$
thus involves $2j$ nodes and our fusion prescription now selects the 
following set of link states
\psset{unit=.48cm}
\setlength{\unitlength}{.48cm}
\bea
\Lc_{\Rc_{j,1}}\ :\hspace{1.3cm} \Bigg\{\;\
   \begin{pspicture}(0,0.63)(6,1.8)
\psarc[linecolor=Maroon,linewidth=1.5pt](2,-0.5){1.25}{0}{37}
\psarc[linecolor=Maroon,linewidth=1.5pt](2,-0.5){1.75}{0}{55}
\psarc[linecolor=Maroon,linewidth=1.5pt](4.5,-0.5){.25}{0}{180}
\psline[linecolor=black,linewidth=1pt,linestyle=dashed,dash=.25 
.25](0,-.5)(0,2.2)
\psline[linecolor=black,linewidth=1pt,linestyle=dashed,dash=.25 
.25](3,-.5)(3,2.2)
\rput(5.05,-0.5){,}
\rput(1.5,-.5){\color{Maroon}$\ldots$}
\rput(3.5,-.8){\tiny$\underbrace{\ }$}
\rput(3.5,-1.3){\tiny $2j\!-\!2$}
\end{pspicture}
   \begin{pspicture}(0,0.63)(5,1.8)
\psarc[linecolor=Maroon,linewidth=1.5pt](2,-0.5){1.25}{0}{37}
\psarc[linecolor=Maroon,linewidth=1.5pt](2,-0.5){1.75}{0}{55}
\psarc[linecolor=Maroon,linewidth=1.5pt](2,-0.5){2.25}{0}{64.5}
\psarc[linecolor=Maroon,linewidth=1.5pt](2,-0.5){2.75}{0}{69.5}
\psline[linecolor=black,linewidth=1pt,linestyle=dashed,dash=.25 
.25](0,-.5)(0,2.2)
\psline[linecolor=black,linewidth=1pt,linestyle=dashed,dash=.25 
.25](3,-.5)(3,2.2)
\rput(1.5,-.5){\color{Maroon}$\ldots$}
\rput(4,-.8){\tiny$\underbrace{\phantom{xxxxxx}}$}
\rput(4,-1.3){\tiny $2j$}
\end{pspicture}
\ \Bigg\}
\label{LR}
\\    \nonumber
\eea

According to the discussion following the decomposition (\ref{LLL}), we can
simply {\em define} the (finitized version of the)
indecomposable representation $\Rc_{j,1}$ by imposing
the boundary condition just introduced. This applies to the 
double-row transfer matrix
$\Db(u)$ as well as to the Hamiltonian limit thereof which can be described
as in (\ref{Hs}) by
\psset{unit=.6cm}
\setlength{\unitlength}{.6cm}
\be
  \Hb_{\Rc_{j,1}}\ =\ \begin{pspicture}[.45](-.25,.5)(10,3.6)
\conn{(0,1)}{(10,3)}
\rput(3.3,1.9){$\Hb_{(1,1)}$}
\psline[linecolor=black,linewidth=1pt,linestyle=dashed,dash=.25 
.25](6,.7)(6,3.3)
\psline[linecolor=blue,linewidth=1.5pt](6.5,1)(6.5,3)
\psline[linecolor=blue,linewidth=1.5pt](7.5,1)(7.5,3)
\psline[linecolor=blue,linewidth=1.5pt](9.5,1)(9.5,3)
\rput(8.57,2){\color{blue}$\dots$}
\end{pspicture}
\label{HsR}
\ee
Here the bulk Hamiltonian has been augmented by the identity $2j$-tangle
and it is recapitulated that the boundary condition dictates that 
$\Hb_{\Rc_{j,1}}$ acts
on the link states (\ref{LR}) with no left-boundary nodes.

As a first example, we consider
\be
  (1,3)\otimes_f{\cal R}_{1,1}\ =\ \Rc_{1,1}\oplus\Rc_{2,1}
\label{13R}
\ee
explicitly for $N=4$. The set of link states decomposes as
\psset{unit=.48cm}
\setlength{\unitlength}{.48cm}
\bea
  \Lc_{(1,3)|\Rc_{1,1}}^{(2)}&=&\Big\{
   \begin{pspicture}(4,2)
\psarc[linecolor=Maroon,linewidth=1.5pt](2,-0.5){1.75}{0}{180}
\psarc[linecolor=Maroon,linewidth=1.5pt](2,-0.5){1.25}{0}{180}
\psarc[linecolor=Maroon,linewidth=1.5pt](1.5,-0.5){.25}{0}{180}
\psarc[linecolor=Maroon,linewidth=1.5pt](2.5,-0.5){.25}{0}{180}
\psline[linecolor=black,linewidth=1pt,linestyle=dashed,dash=.25 
.25](1,-.75)(1,1.5)
\psline[linecolor=black,linewidth=1pt,linestyle=dashed,dash=.25 
.25](3,-.75)(3,1.5)
\rput(4.02,-0.5){,}
\end{pspicture}\
   \begin{pspicture}(4,2)
\psarc[linecolor=Maroon,linewidth=1.5pt](2,-0.5){1.75}{0}{180}
\psarc[linecolor=Maroon,linewidth=1.5pt](2,-0.5){1.25}{0}{180}
\psarc[linecolor=Maroon,linewidth=1.5pt](2,-0.5){.75}{0}{180}
\psarc[linecolor=Maroon,linewidth=1.5pt](2,-0.5){.25}{0}{180}
\psline[linecolor=black,linewidth=1pt,linestyle=dashed,dash=.25 
.25](1,-.75)(1,1.5)
\psline[linecolor=black,linewidth=1pt,linestyle=dashed,dash=.25 
.25](3,-.75)(3,1.5)
\end{pspicture}
  \Big\}  \nn
  \Lc_{(1,3)|\Rc_{1,1}}^{(1)}&=&\Big\{
   \begin{pspicture}(4,2)
\psarc[linecolor=Maroon,linewidth=1.5pt](2,-0.5){1.75}{0}{180}
\psarc[linecolor=Maroon,linewidth=1.5pt](1,-0.5){.25}{0}{180}
\psarc[linecolor=Maroon,linewidth=1.5pt](2,-0.5){.25}{0}{180}
\psarc[linecolor=Maroon,linewidth=1.5pt](3,-0.5){.25}{0}{180}
\psline[linecolor=black,linewidth=1pt,linestyle=dashed,dash=.25 
.25](1,-.75)(1,1.5)
\psline[linecolor=black,linewidth=1pt,linestyle=dashed,dash=.25 
.25](3,-.75)(3,1.5)
\rput(4.02,-0.5){,}
\end{pspicture}\
   \begin{pspicture}(4,2)
\psarc[linecolor=Maroon,linewidth=1.5pt](2,-0.5){1.75}{0}{180}
\psarc[linecolor=Maroon,linewidth=1.5pt](1.5,-0.5){.75}{0}{180}
\psarc[linecolor=Maroon,linewidth=1.5pt](1.5,-0.5){.25}{0}{180}
\psarc[linecolor=Maroon,linewidth=1.5pt](3,-0.5){.25}{0}{180}
\psline[linecolor=black,linewidth=1pt,linestyle=dashed,dash=.25 
.25](1,-.75)(1,1.5)
\psline[linecolor=black,linewidth=1pt,linestyle=dashed,dash=.25 
.25](3,-.75)(3,1.5)
\rput(4.02,-0.5){,}
\end{pspicture}\
   \begin{pspicture}(4,2)
\psarc[linecolor=Maroon,linewidth=1.5pt](2,-0.5){1.75}{0}{180}
\psarc[linecolor=Maroon,linewidth=1.5pt](1,-0.5){.25}{0}{180}
\psarc[linecolor=Maroon,linewidth=1.5pt](2.5,-0.5){.25}{0}{180}
\psarc[linecolor=Maroon,linewidth=1.5pt](2.5,-0.5){.75}{0}{180}
\psline[linecolor=black,linewidth=1pt,linestyle=dashed,dash=.25 
.25](1,-.75)(1,1.5)
\psline[linecolor=black,linewidth=1pt,linestyle=dashed,dash=.25 
.25](3,-.75)(3,1.5)
\end{pspicture}
  \Big\}  \nn
  \Lc_{(1,3)|\Rc_{1,1}}^{(0)}&=&\Big\{
   \begin{pspicture}(4,2)
\psarc[linecolor=Maroon,linewidth=1.5pt](1,-0.5){.75}{0}{180}
\psarc[linecolor=Maroon,linewidth=1.5pt](1,-0.5){.25}{0}{180}
\psarc[linecolor=Maroon,linewidth=1.5pt](3,-0.5){.25}{0}{180}
\psarc[linecolor=Maroon,linewidth=1.5pt](3,-0.5){.75}{0}{180}
\psline[linecolor=black,linewidth=1pt,linestyle=dashed,dash=.25 
.25](1,-.75)(1,1.15)
\psline[linecolor=black,linewidth=1pt,linestyle=dashed,dash=.25 
.25](3,-.75)(3,1.15)
\rput(4.02,-0.5){,}
\end{pspicture}\
   \begin{pspicture}(4,2)
\psarc[linecolor=Maroon,linewidth=1.5pt](1,-0.5){.75}{0}{180}
\psarc[linecolor=Maroon,linewidth=1.5pt](1,-0.5){.25}{0}{180}
\psarc[linecolor=Maroon,linewidth=1.5pt](2.5,-0.5){.25}{0}{180}
\psarc[linecolor=Maroon,linewidth=1.5pt](3.5,-0.5){.25}{0}{180}
\psline[linecolor=black,linewidth=1pt,linestyle=dashed,dash=.25 
.25](1,-.75)(1,1.15)
\psline[linecolor=black,linewidth=1pt,linestyle=dashed,dash=.25 
.25](3,-.75)(3,1.15)
\rput(4.02,-0.5){,}
\end{pspicture}\
   \begin{pspicture}(4,2)
\psarc[linecolor=Maroon,linewidth=1.5pt](1,-0.5){.25}{0}{180}
\psarc[linecolor=Maroon,linewidth=1.5pt](2,-0.5){.25}{0}{180}
\psarc[linecolor=Maroon,linewidth=1.5pt](1.5,-0.5){1.25}{0}{180}
\psarc[linecolor=Maroon,linewidth=1.5pt](3.5,-0.5){.25}{0}{180}
\psline[linecolor=black,linewidth=1pt,linestyle=dashed,dash=.25 
.25](1,-.75)(1,1.15)
\psline[linecolor=black,linewidth=1pt,linestyle=dashed,dash=.25 
.25](3,-.75)(3,1.15)
\rput(4.02,-0.5){,}
\end{pspicture}\
   \begin{pspicture}(4,2)
\psarc[linecolor=Maroon,linewidth=1.5pt](1.5,-0.5){.75}{0}{180}
\psarc[linecolor=Maroon,linewidth=1.5pt](1.5,-0.5){1.25}{0}{180}
\psarc[linecolor=Maroon,linewidth=1.5pt](1.5,-0.5){.25}{0}{180}
\psarc[linecolor=Maroon,linewidth=1.5pt](3.5,-0.5){.25}{0}{180}
\psline[linecolor=black,linewidth=1pt,linestyle=dashed,dash=.25 
.25](1,-.75)(1,1.15)
\psline[linecolor=black,linewidth=1pt,linestyle=dashed,dash=.25 
.25](3,-.75)(3,1.15)
\end{pspicture}
  \Big\}
\eea
Acting on these link states listed in the order indicated here, the 
matrix representation
of the Hamiltonian $\Hb_{(1,3)|\Rc_{1,1}}$ is
\be
  \Hb_{(1,3)|\Rc_{1,1}}\ =\ -\mbox{\scriptsize $\begin{pmatrix}
  0&2&0&1&1&0&0&0&0\\
  1&0&0&0&0&0&0&0&0\\
  0&0&0&1&1&1&0&0&0\\
  0&0&1&0&0&0&0&0&0\\
  0&0&1&0&0&0&0&0&0\\
  0&0&0&0&0&0&0&0&0\\
  0&0&0&0&0&1&0&1&0\\
  0&0&0&0&0&0&1&0&1\\
  0&0&0&0&0&0&0&1&0  \end{pmatrix}$}
\ee
Following the outline above, we work out the finitized dilatation generator
\be
  L_0^{(1,3)|\Rc_{1,1}}\ =\ {\rm diag}\Big[0,\begin{pmatrix} 0&1\\ 
0&0\end{pmatrix},
   1,\begin{pmatrix} 1&1\\ 0&1\end{pmatrix},2,
   \begin{pmatrix} 2&1\\ 0&2\end{pmatrix}\Big]
\ee
as well as those associated to $\Hb_{\Rc_{1,1}}$ and 
$\Hb_{\Rc_{2,1}}$, that is,
\be
  L_0^{\Rc_{1,1}}\ =\ {\rm diag}\Big[\begin{pmatrix} 0&1\\ 0&0\end{pmatrix},
   1,\begin{pmatrix} 2&1\\ 0&2\end{pmatrix}\Big],
  \ \ \ \ \ \ \ L_0^{\Rc_{2,1}}\ =\
  {\rm diag}\Big[0,\begin{pmatrix} 1&1\\ 0&1\end{pmatrix},2\Big]
\label{L01121}
\ee
In accordance with (\ref{13R}), we see that, up to a permutation operation,
\be
  L_0^{(1,3)|\Rc_{1,1}}\ =\ {\rm diag}[L_0^{\Rc_{1,1}},L_0^{\Rc_{2,1}}]
\ee
in which case the finitized partition function decomposes as
\bea
  Z^{(4)}_{(1,3)|\Rc_{1,1}}(q)\ =\ 
\chi^{(4)}_{\Rc_{1,1}}(q)+\chi^{(4)}_{\Rc_{2,1}}(q)
   &=&q^{1/12}[(2+q+2q^2)+(1+2q+q^2)]\nn
   &=&q^{1/12}(3+3q+3q^2)
\eea

As a second example, we consider
\be
  (1,2)\otimes_f{\cal R}_{1,1}\ =\ (1,2)\oplus(1,2)\oplus(1,4)
\label{12R}
\ee
illustrating that fusion of an indecomposable representation does {\em not}
necessarily produce an indecomposable representation.
We demonstrate this explicitly for $N=3$ in which case
\psset{unit=.48cm}
\setlength{\unitlength}{.48cm}
\bea
  {\cal L}_{(1,2)|\Rc_{1,1}}^{(1)}&=&\Big\{
  \begin{pspicture}(3,1.5)
\psarc[linecolor=Maroon,linewidth=1.5pt](1,-0.5){.25}{0}{180}
\psarc[linecolor=Maroon,linewidth=1.5pt](1.5,-0.5){1.25}{0}{180}
\psarc[linecolor=Maroon,linewidth=1.5pt](2,-0.5){.25}{0}{180}
\psline[linecolor=black,linewidth=1pt,linestyle=dashed,dash=.25 
.25](.5,-.75)(.5,1.05)
\psline[linecolor=black,linewidth=1pt,linestyle=dashed,dash=.25 
.25](2,-.75)(2,1.05)
\rput(3.02,-0.5){,}
\end{pspicture}\
\begin{pspicture}(3,1.5)
\psarc[linecolor=Maroon,linewidth=1.5pt](1.5,-0.5){.25}{0}{180}
\psarc[linecolor=Maroon,linewidth=1.5pt](1.5,-0.5){.75}{0}{180}
\psarc[linecolor=Maroon,linewidth=1.5pt](1.5,-0.5){1.25}{0}{180}
\psline[linecolor=black,linewidth=1pt,linestyle=dashed,dash=.25 
.25](.5,-.75)(.5,1.05)
\psline[linecolor=black,linewidth=1pt,linestyle=dashed,dash=.25 
.25](2,-.75)(2,1.05)
\end{pspicture}
      \Big\}\nn
  {\cal L}_{(1,2)|\Rc_{1,1}}^{(0)}&=&\big\{
  \begin{pspicture}(3,1.5)
\psarc[linecolor=Maroon,linewidth=1.5pt](0.5,-0.25){.25}{0}{180}
\psarc[linecolor=Maroon,linewidth=1.5pt](1.5,-0.25){.25}{0}{180}
\psarc[linecolor=Maroon,linewidth=1.5pt](2.5,-0.25){.25}{0}{180}
\psline[linecolor=black,linewidth=1pt,linestyle=dashed,dash=.25 
.25](.5,-.5)(.5,.75)
\psline[linecolor=black,linewidth=1pt,linestyle=dashed,dash=.25 
.25](2,-.5)(2,.75)
\rput(3.02,-0.25){,}
\end{pspicture}\
  \begin{pspicture}(3,1.5)
\psarc[linecolor=Maroon,linewidth=1.5pt](1,-0.25){.25}{0}{180}
\psarc[linecolor=Maroon,linewidth=1.5pt](1,-0.25){.75}{0}{180}
\psarc[linecolor=Maroon,linewidth=1.5pt](2.5,-0.25){.25}{0}{180}
\psline[linecolor=black,linewidth=1pt,linestyle=dashed,dash=.25 
.25](.5,-.5)(.5,.75)
\psline[linecolor=black,linewidth=1pt,linestyle=dashed,dash=.25 
.25](2,-.5)(2,.75)
\rput(3.02,-0.25){,}
\end{pspicture}\
  \begin{pspicture}(3,1.5)
\psarc[linecolor=Maroon,linewidth=1.5pt](0.5,-0.25){.25}{0}{180}
\psarc[linecolor=Maroon,linewidth=1.5pt](2,-0.25){.25}{0}{180}
\psarc[linecolor=Maroon,linewidth=1.5pt](2,-0.25){.75}{0}{180}
\psline[linecolor=black,linewidth=1pt,linestyle=dashed,dash=.25 
.25](.5,-.5)(.5,.75)
\psline[linecolor=black,linewidth=1pt,linestyle=dashed,dash=.25 
.25](2,-.5)(2,.75)
\end{pspicture}
  \big\}
\label{L12R}
\eea
Even though these sets of link states appear the same as those in 
(\ref{L1212}),
the separation into bulk and boundary nodes is different.
Acting on the link states (\ref{L12R}) listed in the order indicated, 
the matrix representation
of the Hamiltonian $\Hb_{(1,2)|\Rc_{1,1}}$ is
\be
  \Hb_{(1,2)|\Rc_{1,1}}\ =\ -\begin{pmatrix}
   0&1&0&0&1\\
   1&0&0&0&0\\
   0&0&0&1&1\\
   0&0&1&0&0\\
   0&0&0&0&0    \end{pmatrix}
\ee
which is seen to be diagonalizable. The associated finitized partition function
decomposes as
\be
  Z^{(3)}_{(1,2)|\Rc_{1,1}}(q)\ =\ 
2\chi^{(3)}_{(1,2)}(q)+\chi^{(3)}_{(1,4)}(q)\ =\
   q^{1/12}[2(1+q^2)+q]\ =\ q^{1/12}(2+q+2q^2)
\ee
in accordance with (\ref{12R}).

As a third example, we consider
\be
  \Rc_{1,1}\otimes_f\Rc_{1,1}\ =\ \Rc_{1,1}\oplus\Rc_{1,1}\oplus\Rc_{2,1}
\label{RR}
\ee
For $N=4$, the numbers of link states are
\be
  |\Lc_{\Rc_{1,1}|\Rc_{1,1}}^{(2)}|\ =\ 2 \ ,\ \ \ \ \ \
  |\Lc_{\Rc_{1,1}|\Rc_{1,1}}^{(1)}|\ =\ 3 \ ,\ \ \ \ \ \
  |\Lc_{\Rc_{1,1}|\Rc_{1,1}}^{(0)}|\ =\ 9
\ee
based on which we have worked out the associated finitized dilatation generator
\be
  L_0^{\Rc_{1,1}|\Rc_{1,1}}\ =\ {\rm diag}\Big[0,\begin{pmatrix} 0&1\\ 
0&0\end{pmatrix},
   \begin{pmatrix} 0&1\\ 0&0\end{pmatrix},1,1,
   \begin{pmatrix} 1&1\\ 0&1\end{pmatrix},2,
   \begin{pmatrix} 2&1\\ 0&2\end{pmatrix},
   \begin{pmatrix} 2&1\\ 0&2\end{pmatrix}\Big]
\ee
The generators $L_0^{\Rc_{1,1}}$ and $L_0^{\Rc_{2,1}}$
for $N=4$ are given in (\ref{L01121}).
In accordance with (\ref{RR}), we find that, up to a permutation operation,
\be
  L_0^{\Rc_{1,1}|\Rc_{1,1}}\ =\ {\rm 
diag}[L_0^{\Rc_{1,1}},L_0^{\Rc_{1,1}},L_0^{\Rc_{2,1}}]
\ee
yielding the following decomposition of the associated finitized 
partition function
\be
  Z^{(4)}_{\Rc_{1,1}|\Rc_{1,1}}(q)\ =\ 
2\chi^{(4)}_{\Rc_{1,1}}(q)+\chi^{(4)}_{\Rc_{2,1}}(q)
\ee

Based on these examples and many others, we conjecture the following 
set of fusion rules.
\\[.3cm]
\noindent {\bf Fusion rules} {\em The fusion algebra of the representations}
$(1,s)$ {\em and} $\Rc_{j,1}$ {\em where} $s,j=1,2,3,\ldots$ {\em is given by}
\bea
  (1,2j_1)\otimes_f(1,2j_2)&=&\bigoplus_j\ \Rc_{j,1}\nn
  (1,2j_1)\otimes_f\Rc_{j_2,1}&=&\bigoplus_j\ 
[(1,2j-2)\oplus(1,2j)\oplus(1,2j)\oplus
    (1,2j+2)]\nn
  \Rc_{j_1,1}\otimes_f\Rc_{j_2,1}&=&\bigoplus_j\ 
[\Rc_{j-1,1}\oplus\Rc_{j,1}\oplus\Rc_{j,1}
    \oplus\Rc_{j+1,1}]\nn
  (1,2j_1-1)\otimes_f(1,2j_2-1)&=&\bigoplus_{j=|j_1-j_2|+1}^{j_1+j_2-1} 
(1,2j-1)   \nn
  (1,2j_1-1)\otimes_f(1,2j_2)&=&\bigoplus_{j=|j_1-j_2-\hf|+\hf}^{j_1+j_2-1}(1,2j) 
\nn
  (1,2j_1-1)\otimes_f\Rc_{j_2,1}&=&\bigoplus_{j=|j_1-j_2-\hf|+\hf}^{j_1+j_2-1} 
\Rc_{j,1}
\label{fusion}
\eea
{\em where} $(1,0)=\Rc_{0,1}=0$. {\em The three unspecified sums are over}
\be
  j\ =\ |j_1-j_2|+1,|j_1-j_2|+3,\ldots,j_1+j_2-1
\ee
{\em whereas the three specified sums are with} unit {\em increments.}\\[.3cm]
As expected, $(1,1)$ is seen to play the role of the identity representation.
It is noted that the set of representations $(1,s)$ and $\Rc_{j,1}$
is closed under fusion. Several distinct fusion subalgebras and 
ideals are easily identified.
It is also noted that the fourth and fifth fusion rules are identical 
to (\ref{sl2}) with
$s_1$ odd.

We stress that we have considered fusion generated from
the $(1,s)$ representations and that they are only irreducible for 
$s=1$ or $s$ even.
The fusion rules in the CFT approach~\cite{GK}, on the other hand,
are expressed in terms of the irreducible representations $\Vc_{j,1}$ 
and $\Vc_{j,2}$ only and the indecomposable representations resulting 
from fusion. A discussion of
the irreducible representations from the {\em lattice} point of view 
will appear elsewhere.

Consistent with the fusion rules of~\cite{GK}, we have observed that 
our fusion rules (\ref{fusion}) are reproduced by a simple extension 
by linearity, namely
\be
  (1,2j)\ \sim\ \Vc_{j,2},\qquad (1,2j-1)\ \sim\ 
\Vc_{j-1,1}+\Vc_{j,1},\qquad j=1,2,\ldots
\ee
where $\Vc_{0,1}=0$. We have already identified the indecomposable 
representations $\Rc_{j,1}$
with those appearing in~\cite{GK}.

\section{Conclusion}

In this paper, we have solved exactly a model of critical dense 
polymers on strips of {\it finite} width. This has been achieved, without invoking 
an $n\to 0$ or $Q\to 0$ limit, by solving directly our model of critical dense 
polymers within the framework of Yang-Baxter integrability.
The calculations have been carried out 
for an infinite family of $(1,s)$ integrable boundary conditions 
which impose $\ell=s-1$ defects in the bulk.
Our study of the physical combinatorics, which is the classification 
of the physical states using combinatorial objects, has revealed some 
interesting connections to $q$-Narayana numbers
and natural generalizations thereof.
Using the planar TL algebra, Yang-Baxter techniques 
and functional equations in the form of inversion identities, 
we have been able to calculate analytically the bulk and boundary 
free energies, the central charge $c=-2$, the conformal weights 
$\Delta_s=\frac{(2-s)^2-1}{8}$ and conformal characters in accord  
with previous results~\cite{Saleur87,Duplantier,SaleurSUSY,ReSa01}.
More particularly, since we were able to solve the model exactly for finite system sizes, 
we have explicitly confirmed the {\it finitized} 
conformal characters proposed in \cite{PRZ}.

In support of our claim that the CFT associated with critical dense 
polymers is logarithmic, we have shown in explicit examples how 
fusion of our $(1,s)$ boundary conditions in some cases lead to 
indecomposable representations ${\cal R}_{j,1}$. We have argued, by 
implementing fusion diagrammatically, that fusion among the $(1,s)$ 
representations and indecomposable representations ${\cal R}_{j,1}$ 
closes, and we have conjectured the general form of the associated 
fusion rules. A proper comparison with the fusion algebra of 
Gaberdiel and Kausch~\cite{GK}, however, requires that our fusion 
algebra is formulated in terms of {\em irreducible} representations 
alongside the indecomposable representations. A discussion of these 
irreducible representations
from the lattice point of view will appear elsewhere. We have 
nevertheless found that
the fusion rules presented
here are in accordance with those in~\cite{GK} in the sense that our 
fusion rules can be obtained from theirs by extending by linearity to 
the $(1,s)$ representations.

Again, it is a well-known abstract result~\cite{MartinBook} that the TL algebra admits indecomposable representations at roots of unity. We emphasize that our motivation here is to construct explicit lattice Hamiltonians for these representations in the case of critical dense polymers, to relate them to physical boundary conditions and to understand their fusion properties in physical terms. 

To claim a complete understanding of our model of critical dense 
polymers, we should
study its extension from the strip to other topologies.
Most importantly, the model should be examined on the geometric 
cylinder and torus where the effects of non-contractible loops will 
play a role. This constitutes work in progress.
\vskip.5cm
\noindent{\em Acknowledgments}
\vskip.1cm
\noindent
This work is supported by the Australian Research Council.
The authors thank Jean-Bernard Zuber for discussions.

\end{document}